%% file: main.tex
\documentclass[sigplan,nonacm]{acmart}

\include{packages}
\include{macros}

\begin{document}
\sloppy

\author{Thomas Gassmann}
\affiliation{%
  \institution{ETH Zurich}
  \country{Switzerland}
}

\author{Stefanos Chaliasos}
\affiliation{%
  \institution{UCL Centre for Blockchain Technologies}
  \country{United Kingdom}
}
\affiliation{%
  \institution{zkSecurity}
  \country{USA}
}

\author{Thodoris Sotiropoulos}
\affiliation{%
  \institution{ETH Zurich}
  \country{Switzerland}
}

\author{Zhendong Su}
\affiliation{%
  \institution{ETH Zurich}
  \country{Switzerland}
}

\title[Evaluating Compiler Optimization Impacts on zkVM Performance]
      {Evaluating Compiler Optimization Impacts on\\ zkVM Performance}

\begin{abstract}

Zero-knowledge proofs (ZKPs) 
are the cornerstone of programmable cryptography.
They enable (1) privacy-preserving 
and verifiable computation 
across blockchains,
and (2) an expanding range of off-chain applications 
such as credential schemes. 
Zero-knowledge virtual machines (zkVMs) 
lower the barrier by turning ZKPs 
into a drop-in backend 
for standard compilation pipelines.
This lets developers write proof-generating programs in
conventional languages
(e.g., Rust or C++)
instead of hand-crafting arithmetic circuits.
However,
these VMs inherit compiler infrastructures
tuned for traditional architectures
rather than for proof systems.
In particular,
standard compiler optimizations assume features
that are absent in zkVMs,
including cache locality,
branch prediction,
or instruction-level parallelism.
Therefore,
their impact on proof generation
is questionable.

We present the first systematic study of
the impact of compiler optimizations on zkVMs.
We evaluate 64 LLVM passes,
six standard optimization levels,
and an unoptimized baseline across 58 benchmarks
on two RISC-V–based zkVMs
(RISC Zero and SP1).
While standard LLVM optimization levels do improve zkVM performance
(over 40\%),
their impact is far smaller than on traditional CPUs,
since their decisions rely on hardware features
rather than proof constraints.
Guided by a performance impact analysis,
we~\emph{slightly} refine a small set of LLVM passes to be zkVM-aware,
improving zkVM execution time
by up to~\empirical{45\%}
(average +\empirical{4.6\%} on RISC Zero,
+\empirical{1\%} on SP1)
and achieving consistent proving-time gains.
Our work highlights
the potential of compiler-level optimizations for zkVM performance
and opens new directions for zkVM-specific passes,
backends,
and superoptimizers.
\end{abstract}

\maketitle 

\section{Introduction}
\label{sec:intro}
Zero-knowledge proofs (ZKPs)~\cite{zkp} 
let a prover 
convince a verifier 
that a statement holds, 
optionally, 
without revealing anything beyond its truth. 
After decades as a theoretical concept, 
advances in proof systems 
have made ZKPs practical at scale~\cite{pinocchio,groth16,plonk}. 
Still, two obstacles have thwarted adoption: 
(1) developers had to handcraft low-level arithmetic circuits~\cite{circom}, 
which is both hard and error-prone~\cite{sokzkps}, 
and (2) proving remains costly despite recent progress~\cite{zkbench}.

Zero-knowledge virtual machines (zkVMs)~\cite{risc0,sp1,jolt,Goldberg2021CairoA} 
address both issues by enabling
developers to reuse well-established languages and toolchains.
This is achieved
by compiling ordinary programs (e.g., Rust) 
to a standard ISA such as RISC-V~\cite{riscv}, 
then emulating the binary 
to record an execution trace 
that a zkVM back-end proves. 
This design leverages tools such as LLVM~\cite{llvm}, 
avoids manually-written circuits, 
and benefits from techniques 
such as GPU proving~\cite{gzkp,sp1-hardware-acceleration}, 
recursion~\cite{nova,risc0-recursion}, 
and precompiles for heavy primitives. 
zkVMs are already deployed 
in key blockchain applications~\cite{sp1-bridge,risc0-rollup},
and are increasingly adopted in off-chain systems,
such as anonymous credentials 
and client-side proving~\cite{ligero}.
A prime example is
Ethereum's long-term roadmap
which is heavily centered around zkVMs.
Current proposals include
replacing the EVM with RISC-V to accelerate proving~\cite{simplel1},
advancing formal-verification efforts to harden zkVM correctness~\cite{verifiedzkevm}, 
and initiatives such as ETHProofs that monitor progress across both fronts~\cite{ethproofs}.

Despite this progress, 
performance remains the central hurdle
and a competitive factor among
zkVM vendors~\cite{sp1-benchmarks,ethproofs}.
zkVMs' architecture adopts an execution model fundamentally
different from traditional CPUs,
where every instruction becomes a set of constraints to be proved.
Yet,
compilers still optimize for hardware features
that are not relevant to zkVMs,
such as caches,
out-of-order execution,
and instruction-level parallelism.
All these raise the following question:
\textit{How do compiler-level techniques
(i.e., optimizations) influence zkVM performance,
and to what extent?}

This open question motivates this study,
which aims to investigate how existing optimizations
influence zkVM performance
and identify concrete opportunities for improvement.
We seek answers for the following three research questions:

\begin{enumerate}[label={\bf RQ\arabic*}, leftmargin=2.6\parindent]
\item How do standard compiler optimizations impact zkVM performance 
when applied individually?
(Section~\ref{sec:individual-passes})

\item Are there combinations of optimizations 
that yield significant performance improvements 
or degradations?
(Section~\ref{sec:combinations-optimizations})

\item
How does the impact of optimizations on zkVMs 
compare to traditional architectures? 
(Section~\ref{sec:x86-comparison})

\end{enumerate}

To answer these research questions, 
we benchmark 58 programs on two zkVMs, 
RISC Zero~\cite{risc0} 
and SP1~\cite{sp1}, 
across 71 optimization profiles
that include~\tpasses\ individual LLVM passes,
six default optimization levels (e.g., {\tt -O3}),
and an unoptimized baseline.
The selected zkVMs are
the most mature and widely used ones,
targeting a conventional ISA
that leverages well-established toolchains~\cite{risc0-github,sp1-github},
such as LLVM.
Our benchmarks cover a wide range of programs, 
including cryptographic primitives 
and mathematical computations. 
The effects of optimizations are quantified
through three metrics:
cycle count,
zkVM execution time,
and proving time.

\point{Contributions}
We make the following contributions:

\begin{itemize}
\item
We present the first systematic study of
how traditional compiler optimizations
(both individual passes
and their combinations)
affect zkVM performance,
and compare the results to traditional CPUs
(Section~\ref{sec:results}).

\item
We identify and empirically validate
key cost components that drive zkVM performance,
that is,
dynamic instruction count,
and paging cycles.
(Section~\ref{sec:cost-components}).

\item
We analyze selected optimizations
based on our cost components,
and derive four optimization principles,
which we validate empirically.
We then apply these principles to refine
existing passes for better zkVM performance.
(Sections~\ref{sec:case-studies} and~\ref{sec:zkvm-specific-optimizations}).

\item
We enumerate the implications of our findings,
and discuss potential future directions on
improving zkVM performance
(Section~\ref{sec:discussion}).
\end{itemize}

\point{Summary of findings}
Some of our representative findings are:
(1) default optimization levels
(e.g., {\tt -O3})
improve both zkVM execution and proving time
by over~\empirical{40\%};
(2) however,
these gains are significantly smaller than
on traditional CPUs,
as many individual passes exhibit opposing effects,
mainly due to wrong decision heuristics
for inlining,
or loop unrolling.
These heuristics rely on
exploiting hardware-specific features
(e.g., cache locality)
rather than minimizing the number of the executed instructions;
and (3) autotuning proves to be an effective approach for
performance-critical zkVM programs
(up to~$2.2\times$ speedup compared to {\tt -O3}).

\point{Implications}
We identify root causes of inefficiencies
in certain passes and implement~\emph{slight} modifications to
an existing compiler toolchain (i.e., LLVM).
The {\tt -O3} level in our modified LLVM outperforms
the original LLVM’s {\tt -O3} in most benchmarks,
achieving up to~\empirical{45\%} faster zkVM execution
(avg. +\empirical{4\%} on RISC Zero;
+\empirical{1\%} on SP1)
while also delivering consistent improvements in proving time.
Since zkVM execution and proving are
orders of magnitude more expensive
than execution on traditional CPUs,
even small percentage improvements can translate into
minutes of savings
(Section~\ref{sec:background}).


\section{Background}
\label{sec:background}
\point{Zero-Knowledge Proofs and zkVMs}
A ZKP~\cite{zkp} is a cryptographic primitive 
that allows a prover to convince a verifier 
that a statement is valid,
optionally, without revealing anything beyond its truth. 
A ZKP provides \emph{completeness},
\emph{soundness}, 
and optionally \emph{zero knowledge}.
In the past decade, 
adoption of ZKPs have accelerated with SNARKs~\cite{pinocchio,groth16}, 
which provide non-interactive, 
succinct proofs with fast verification 
and practical proving costs. 
ZKPs have enabled many novel applications~\cite{doyouneed,zksurvey}
such as private transactions~\cite{zcash}
and credential systems~\cite{zklogin,longfellow}.

\begin{figure}[t]
    \centering
    \includegraphics[width=0.48\textwidth]{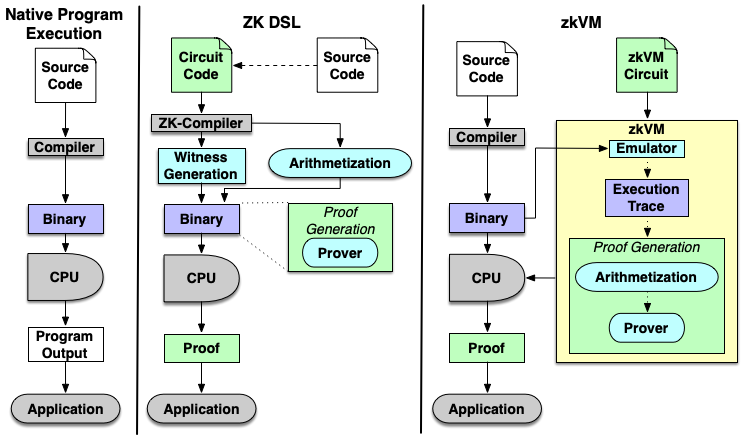}
    \caption{Architecture comparison between native execution, ZK DSLs, and zkVMs.}
    \label{fig:zkvm-stack}
\end{figure}

SNARKs target \emph{circuit satisfiability}~\cite{pinocchio}, 
representing statements as~\emph{arithmetic circuits}. 
Early ZK development used low-level ZK DSLs~\cite{circom},
where a compiler \emph{arithmetizes} the circuit 
into a concrete constraint system 
(e.g., PLONK-style~\cite{plonk}) 
and produces a \emph{witness generator}.
Then,
the \emph{prover} uses the constraints and witness 
to~\emph{produce a proof.} 
Writing low-level ZK circuits is not only error-prone~\cite{sokzkps}, 
but it also requires significant expertise. 
\emph{Zero-knowledge virtual machines} (zkVMs)~\cite{SNARK-vonNeumann} 
raise the abstraction 
by compiling source code
to a target ISA (often RISC-V~\cite{riscv})~\cite{sp1,risc0}.
Then,
the zkVM's \emph{emulator}
executes the binary 
and records an \emph{execution trace} 
in the proper arithmetization, 
and the back-end produces a proof. 
Developers use well-established compilation toolchains
(e.g., LLVM), 
and the zkVM exposes \emph{precompiles} 
for heavy primitives (e.g., SHA-2)
to replace thousands of instructions 
with optimized circuits. 
To make zkVMs practical for computation-heavy applications
(e.g., ZK-Rollups~\cite{WhiteHat_roll_up_token}),
much of the literature focuses on improving
both execution 
and proof generation~\cite{jolt,lasso,Goldberg2021CairoA,sp1,risc0}.
Figure~\ref{fig:zkvm-stack} contrasts native execution, 
ZK DSLs, 
and zkVMs.
For an in-depth introduction of ZKPs, 
we refer the reader to~\cite{thaler2022proofs}.

\point{zkVMs vs.\ traditional architectures}
When generating the execution trace 
and proving it on a zkVM, 
the execution model differs significantly 
from traditional CPU architectures, 
such as x86 or ARM.
Most instructions have near-uniform cost, 
and there is no instruction latency 
due to cache hits/misses 
and other micro-architectural effects.
Memory access is relatively cheaper on zkVMs
(cache misses are not penalized), 
but page-ins and page-outs can be significantly more expensive.
Further, 
there is no parallelism or multithreading, 
and floating-point operations 
are not supported natively, 
so they need to be emulated, 
which makes them extremely costly. 
We detail a full list of differences in Appendix~\ref{app:zkvm-vs-cpu}.
Notably,
zkVM execution and proving are orders of magnitude slower
than native execution
(milliseconds vs.\ seconds to hours,
cf.\ Figure~\ref{fig:zkvm-costs} Appendix~\ref{app:zkvm-vs-cpu}).

\point{Compiler optimizations and zkVMs}
Compiler optimizations are semantics-preserving transformations 
that aim to improve speed, 
reduce size, 
or enhance other attributes of code~\cite{compiler-engineering,compiler-dragonbook}.
We present two motivating examples 
where optimizations that aid conventional CPUs 
degrade zkVM performance, 
exposing a mismatch between hardware-oriented assumptions 
and the zkVMs' proof-centric cost model.

\textit{Strength reduction.}
LLVM's \emph{strength reduction}
replaces division instruction 
with a shift-and-add sequence 
(Figure~\ref{fig:example_instruction_selection}).
On conventional CPUs, 
this is beneficial 
because division instruction 
is often costlier 
than the equivalent 
bitwise instructions~\cite{instruction-tables-traditional-architectures}.
Our x86 measurements 
show that the transformed code is~$3.5\times$ faster. 
On zkVMs, 
the effect reverses, 
notably proving is 40\% slower in RISC Zero~\cite{risc0}.
The reason is the cost model mismatch: 
on zkVMs,
every operation contributes to the number of constraints in the proof,
and all instructions have roughly uniform cost.
Replacing a single division with multiple bitwise operations
therefore increases the constraint count,
making proving slower.

\begin{figure}[t]
\centering
  \begin{subfigure}[t]{0.49\textwidth}
    \centering
    \begin{minipage}[t]{0.48\textwidth}
\begin{verbatim}
// unoptimized version
example::div:
        li t0, 8
        div a0, a0, t0
        ret
  \end{verbatim}
  \end{minipage}
  \hfill
  \begin{minipage}[t]{0.45\textwidth}
  \begin{verbatim}
// optimized version 
example::div:
        srai a1, a0, 31
        srli a1, a1, 29
        add a0, a0, a1
        srai a0, a0, 3
        ret
  \end{verbatim}
  \end{minipage}
  \caption{
CPU-oriented instruction selection can hurt zkVMs: 
on RISC Zero, the `optimized' form takes 8 cycles vs.\ 4 unoptimized}
  \label{fig:example_instruction_selection}
\end{subfigure}
\hfill
\begin{subfigure}[t]{0.49\textwidth}
  \centering
  \begin{minipage}[t]{0.47\textwidth}
  \begin{verbatim}
/* unoptimized version */
int i, a[N], b[N];
for (i = 0; i < N; i++) {
    a[i] = 1;
    b[i] = 2;
}
  \end{verbatim}
  \end{minipage}
  \hfill
  \begin{minipage}[t]{0.47\textwidth}
  \begin{verbatim}
/* optimized version */
int i, a[N], b[N];
for (i = 0; i < N; i++) {
    a[i] = 1;
}
for (i = 0; i < N; i++) {
    b[i] = 2;
}
  \end{verbatim}
  \end{minipage}
  \caption{
\emph{Loop fission} often helps CPUs 
but can hurt zkVMs by duplicating loop-control work.
$N$ is set to $1048576$.
}
  \label{fig:example_loop_fission}
\end{subfigure}
\caption{Optimizations with negative effects on zkVMs.}
\label{fig:negative_examples}
\end{figure}

\textit{Loop fission}
is an optimization that splits a loop 
into multiple ones 
to enhance locality 
and reduce cache misses 
(Figure~\ref{fig:example_loop_fission}).
On x86,
the transformed code is $\sim$8\% faster.
On zkVMs, 
the effect flips:
duplicating loop control increases proof constraints, 
raising proving time by 5\% on SP1~\cite{sp1}. 
This underscores that cache-centric assumptions 
do not transfer to zkVMs' model.

\section{Methodology}
\label{sec:methodology}

\point{Selection of zkVMs} 
We evaluate two zkVMs 
targeting RISC-V~\cite{riscv} (SP1~\cite{sp1} and RISC Zero~\cite{risc0}), 
chosen for maturity and widespread use in production. 
We also considered zkVMs with non-standard ZK-oriented ISAs such 
as Cairo~\cite{Goldberg2021CairoA} and Valida~\cite{valida}, 
but these stacks bypass existing compiler infrastructures
and rely on less stable transpilers. 

The selected zkVMs can directly run C and Rust programs
that are lowered to RISC-V via
an existing compiler toolchain,
in particular LLVM~\cite{llvm}.
Therefore,
our work specifically investigates the impact of traditional
compiler optimizations~\emph{implemented in LLVM} on zkVMs.

\subsection{Metrics}
We evaluate the impact of optimizations on zkVMs
using three zkVM metrics: 
\emph{cycle count}, 
\emph{zkVM execution time}, 
and \emph{proving time}. 
\emph{Cycle count} 
is the total cycles of the program run on the zkVM
(also known as the~\emph{guest} program),
and is analogous to CPU cycles. 
It captures only the cost of program execution and proving,
and does not include other one-time setup
or initialization overheads,
such as prover initialization.
\emph{zkVM execution time} 
is the wall-clock time 
for the executor to replay the guest 
and produce the full execution trace.
\emph{Proving time} 
is the wall-clock time for the prover 
to generate a proof 
and is typically larger than execution time. 
For native runs (RQ3),
we measure native execution time 
to contrast optimization effects 
on zkVMs and traditional architectures.

\subsection{Benchmark Selection}
We combine standard benchmarks 
along with zkVM-focused workloads 
to cover both large, 
optimization-sensitive programs 
and common zkVM payloads. 
In total, 
our benchmark suite consists of 58 programs 
written in Rust and C/C++.
We use PolyBench~\cite{polybench-c,polybench-rust}, 
NPB~\cite{npb-c,npb-rust}, 
and SPEC CPU 2017~\cite{spec-cpu}, 
which include compute 
and memory-intensive programs. 
From industry benchmark suites, 
we include the a16z crypto zkVM,
the Succinct Labs, 
and the RSP benchmarks~\cite{a16z-benchmarks,succinct-zkvm-perf,succinct-reth}.
These capture typical cryptography-heavy workloads. 
We further add targeted programs,
such as \texttt{regex-match}, 
\texttt{sha256}, 
and \texttt{loop-sum}, 
to exercise loop-heavy, 
memory-heavy, 
and math/crypto patterns frequently seen in zkVMs.
For the complete list of the benchmarks
see Appendix~\ref{app:benchmarks}.

\subsection{Analyzing Optimizations}
\label{sec:research-questions}
To measure the impact of individual optimizations 
as well as standard optimization levels, 
we evaluate a total of 71 optimization profiles, 
which consist of 64 individual LLVM optimization passes, 
1 unoptimized \textit{baseline} profile,
and 6 preset optimization levels 
(\texttt{-O0}, \texttt{-O1}, \texttt{-Oz}, etc.).
For all 64 individual optimization passes 
and the \texttt{baseline} profile,
we also disable Rust's MIR optimizations 
by setting \texttt{mir-opt-level} to \texttt{0}. 
The \texttt{baseline} profile has no optimizations applied 
and serves as a reference point 
for other optimization profiles.

To answer the RQs
outlined in Section~\ref{sec:intro}, 
we first measure each pass 
in isolation on RISC Zero and SP1 
across our benchmark suite, 
collecting cycle count, 
executor time, 
and proving time 
(Section~\ref{sec:individual-passes}) and
quantify the performance losses/gains
compared to the baseline.
Next,
we explore optimization combinations
via genetic autotuning with OpenTuner~\cite{opentuner} 
(Section~\ref{sec:combinations-optimizations}).
Then,
for RQ3,
we run the same optimization profiles 
on traditional CPU (i.e., x86) 
and compare the speedups with those obtained from zkVMs
to pinpoint opportunities for improvement 
and understand which optimizations
are worth revisiting or
redesigning in the zkVM context
(Section~\ref{sec:x86-comparison}).

\point{Setup}
x86 and zkVM \emph{execution} experiments 
run on an AMD EPYC 7742 
with 500GB of RAM 
on Ubuntu 22.04.5. 
\emph{Proving} 
runs on an AMD EPYC 7542 
with 32GB RAM 
on Ubuntu 24.04.2, 
equipped with an NVIDIA GeForce RTX 4090 (24GB VRAM). 
The SP1 prover is executed locally
via its RPC API.
We use SP1 4.1.4 (moongate-server 4.1.0), 
RISC Zero 1.2.4, 
Clang 19.1.7, 
LLVM 19.1.7, 
and Rust 1.85.0.
For both zkVMs we used their default STARK-based proof systems.



\subsection{Threats to Validity}
\label{sec:threats-to-validity}
\point{Internal validity} 
Proving time on SP1 exhibits higher variance than on RISC Zero, 
partly due to the closed-source prover accessed via RPC 
and the associated network overhead. 
To mitigate this,
We use at least 10 samples 
per configuration 
and increase the sample size for noisy cases. 

\point{External validity} 
Our proving time measurements are accelerated on an NVIDIA RTX 4090.
However,
different GPUs or CPUs may change effects. 
We study two popular RISC-V zkVMs, 
but other zkVMs or designs 
(e.g., different ISAs) 
may behave differently. 
Extending the study to additional zkVMs 
is left for future work.
Our RQ3 comparison is limited to x86.
While ARM or RISC-V CPUs could provide additional insights,
we choose x86 since it is currently
the most widely used architecture
for running zkVM workloads.
Finally,
results may not generalize to all workloads,
but we mitigate this by selecting benchmarks spanning scientific,
loop-intensive,
and cryptographic domains.

\point{Construct validity} 
Several benchmarks use reduced input sizes 
to keep proving feasible. 
We evaluate~\tpasses individual LLVM passes 
essentially in isolation with default parameters. 
Many additional passes 
and tune strategies exist, 
and some benefits only appear in specific combinations. 
Although we explored combinations via autotuning, 
the whole phase-ordering space remains intractable.

\section{Results}
\label{sec:results}
\begin{figure*}[t]
\centering
\includegraphics[width=\linewidth]{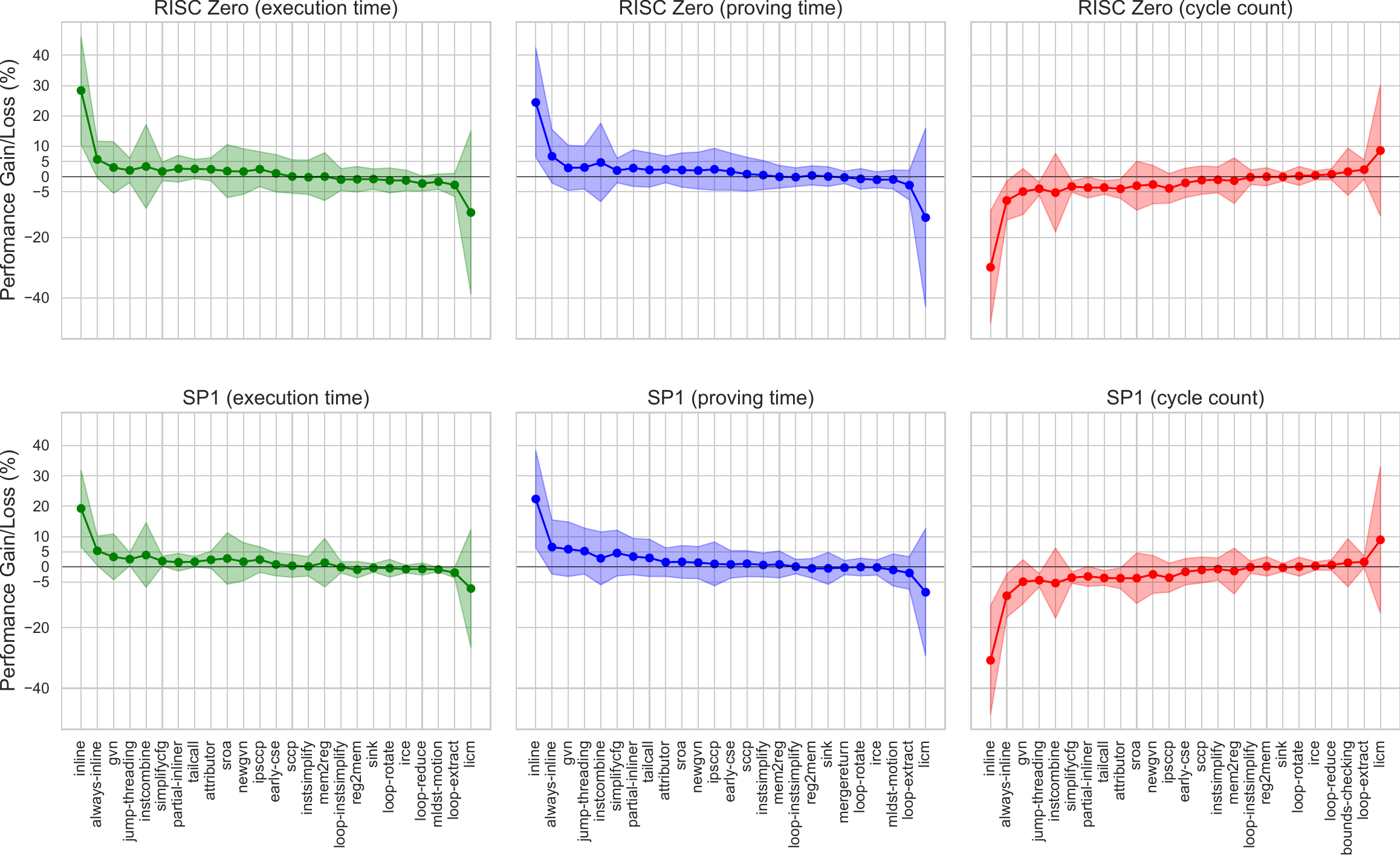}
\caption{The impact of top 25 individual LLVM passes across
all benchmarks,
zkVMs (RISC Zero, SP1),
and metrics (zkVM execution time,
proving time, cycle count).
Lines represent average impact;
shaded areas show standard deviation.
Higher values indicate better performance
for proving time and zkVM execution time,
while lower is better for cycle count.}
\label{fig:individual-impact}
\end{figure*}

In this section,
we present the results for each research question
outlined in Section~\ref{sec:intro}.

\subsection{RQ1: Impact of Individual LLVM Passes}
\label{sec:individual-passes}

Figure~\ref{fig:individual-impact} presents
the top 25 LLVM passes with the highest average impact
(either positive or negative)
across all benchmarks,
zkVMs,
and metrics.
Each result reflects the performance difference relative
to an unoptimized baseline with no passes applied.
The remaining~\empirical{39} passes
have minimal impact
and are omitted for brevity,
as they change execution time
and proving time by only~\empirical{0.9\%}--\empirical{1}\%,
on average.

\point{zkVM execution time}
Only a few passes,
on average,
improve zkVM execution time when applied individually.
Out of~\tpasses passes,
\empirical{12} reduce execution time on RISC Zero
by at least 1\% on average,
and~\empirical{13} do so on SP1.
The most harmful pass on both zkVMs is \texttt{licm},
which increases execution time
by~\empirical{11.8}\% on RISC Zero,
and~\empirical{7.1}\% on SP1.
Section~\ref{sec:analysis-selected-passes}
analyzes the root causes of
this regression in detail.

Inlining is beneficial for both zkVMs:
the {\tt inline} and {\tt always-inline} passes
lead to performance gains of~\empirical{28.4\%}
and~\empirical{5.7\%} on RISC Zero,
and~\empirical{19.3\%} and~\empirical{5.3\%} on SP1,
respectively.
Other passes,
such as {\tt instcombine} or {\tt sroa},
perform well on some benchmarks,
while degrading performance on others.
Both zkVMs exhibit similar trends,
with no pass consistently having opposite effects on RISC Zero and SP1.

\point{Proving time}
Proving time closely follows
the trends observed in zkVM execution.
The most beneficial pass is again~\texttt{inline}
(\empirical{22.4}\% performance gain
on both SP1 and RISC Zero).
The most harmful pass remains~\texttt{licm},
which increases proving time by~\empirical{8.4}\% on SP1
and~\empirical{13.5}\% on RISC Zero.

\point{Cycle count}
Similar trends are observed when
examining cycle count.
Inlining reduces cycle count
with a~\empirical{30.8\%} reduction on SP1 
and a~\empirical{29.9\%} on RISC Zero.
The pass that increases the cycle count the most is {\tt licm},
which explains its negative effect
on zkVM execution and proving.


\begin{figure*}[t]
\centering
\includegraphics[width=0.87\linewidth]{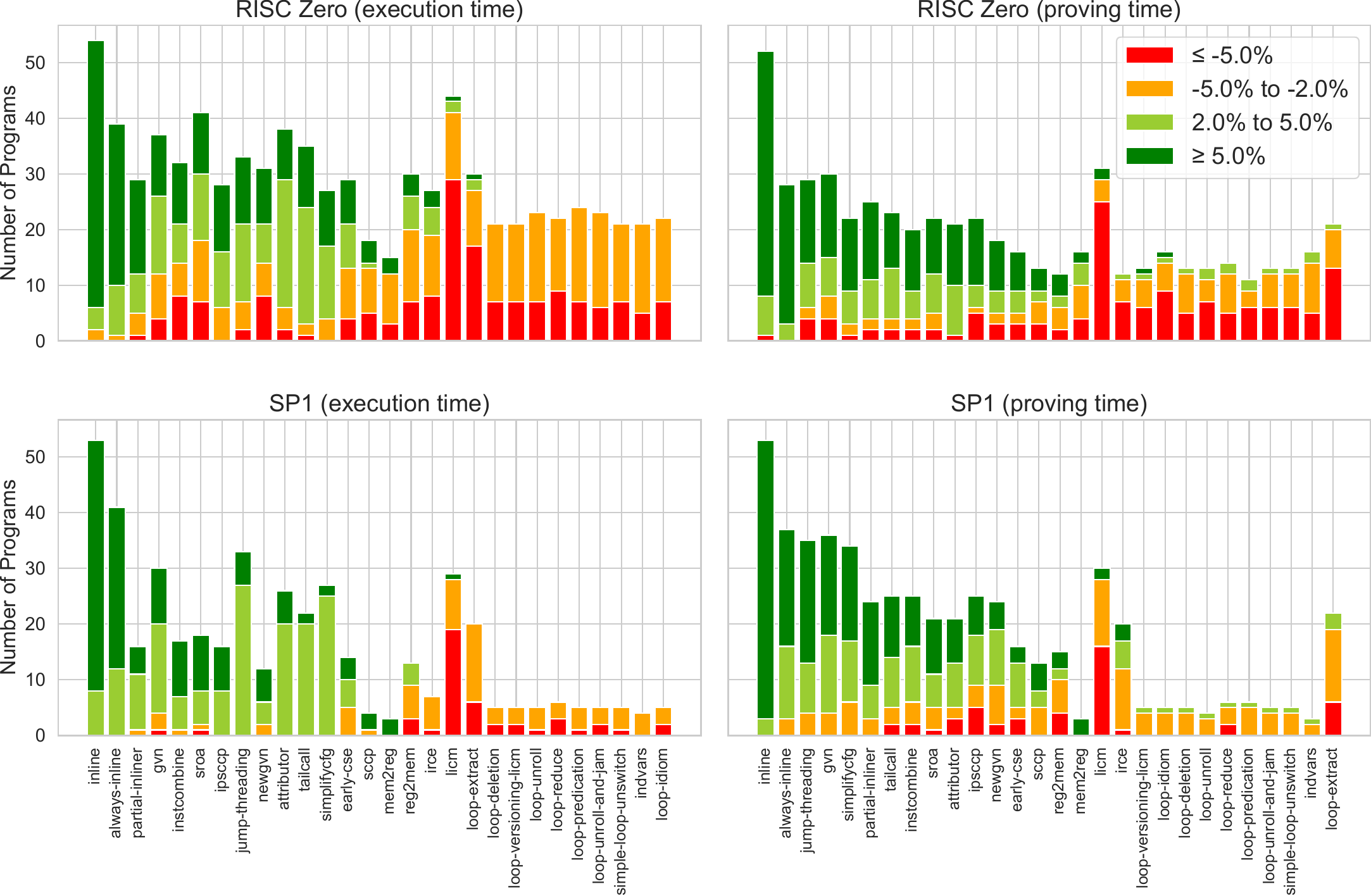}
\caption{
Counting the cases where each pass
leads to (1) severe performance gain ($\ge 5$\%),
moderate performance gain (between $2\%$ and $5\%$),
moderate performance loss (between -2\% and -5\%),
and severe performance loss ($\le -5$\%).
}
\label{fig:prog-impact}
\end{figure*}

\begin{table}[t]
\caption{Number of instances
where optimized code leads to performance gains or losses
in execution and proving time.}
\label{tab:count-impact-zkvm}
\resizebox{0.8\linewidth}{!}{
\begin{tabular}{l|rr|rr}
\toprule
& \multicolumn{2}{c|}{\textbf{Execution}} & \multicolumn{2}{c}{\textbf{Proving}} \bigstrut \\
\midrule
\textbf{zkVM} & \textbf{Gain (\textgreater 2\%)} & \textbf{Loss (\textless -2\%)} & \textbf{Gain (\textgreater 2\%)} & \textbf{Loss (\textless -2\%)} \bigstrut \\
\midrule
RISC Zero     & 370 & 437 & 302 & 241 \\
SP1           & 314 & 124 & 347 & 174 \\                   
\bottomrule
\end{tabular}}
\end{table}

\point{Categories of effects}
For completeness,
Figure~\ref{fig:prog-impact} shows
the number of benchmarks
where each pass causes:
(1) moderate performance loss
(-5\% to -2\%),
(2) severe loss ($\le -5\%$),
(3) moderate gain (2\% to 5\%),
or (4) severe gain ($\ge 5\%$).
Inlining consistently improves
both execution and proving times,
though a few benchmarks still see execution slowdowns
(more details in Section~\ref{sec:analysis-selected-passes}).
In contrast,
loop optimizations
(e.g., {\tt loop-extract},
{\tt licm},
{\tt loop-deletion},
{\tt loop-unroll}) often harm zkVM performance,
especially on RISC Zero.
This is because
loop optimizations in LLVM operate
in Loop Closed SSA (LCSSA) form,
where values defined in a loop
but used outside it are routed through {\tt phi} nodes at the loop exit.
This transformation often requires recomputing addresses
(e.g., via {\tt getelementptr}) or
adding load/store instructions,
especially when iterating over arrays,
leading to extra memory operations that
harm zkVM performance (Section~\ref{sec:analysis-selected-passes}).
Finally,
some passes,
such as {\tt instcombine} on RISC Zero execution,
show a balanced impact,
with similar numbers of benchmarks improving and regressing.
This suggests that certain rewrites in {\tt instcombine} may
hurt zkVM performance
and should be adjusted to better
align with zkVMs' model
(Figure~\ref{fig:example_instruction_selection}).

\point{Differences in zkVMs}
Differences between zkVMs
are evident in Table~\ref{tab:count-impact-zkvm},
which shows how often optimized code results in performance gains
or losses in execution
and proving time for each zkVM.
Overall,
individual optimizations appear to have
a more detrimental effect on RISC Zero,
especially in execution,
where degradation occurs roughly $\empirical{4}\times$
more often compared to SP1.
This could stem from
differences in IRs
and constraint-level optimizations across zkVM.


\subsection{RQ2: Combinations of Optimizations}
\label{sec:combinations-optimizations}

\begin{figure}[t]
\includegraphics[width=1\linewidth]{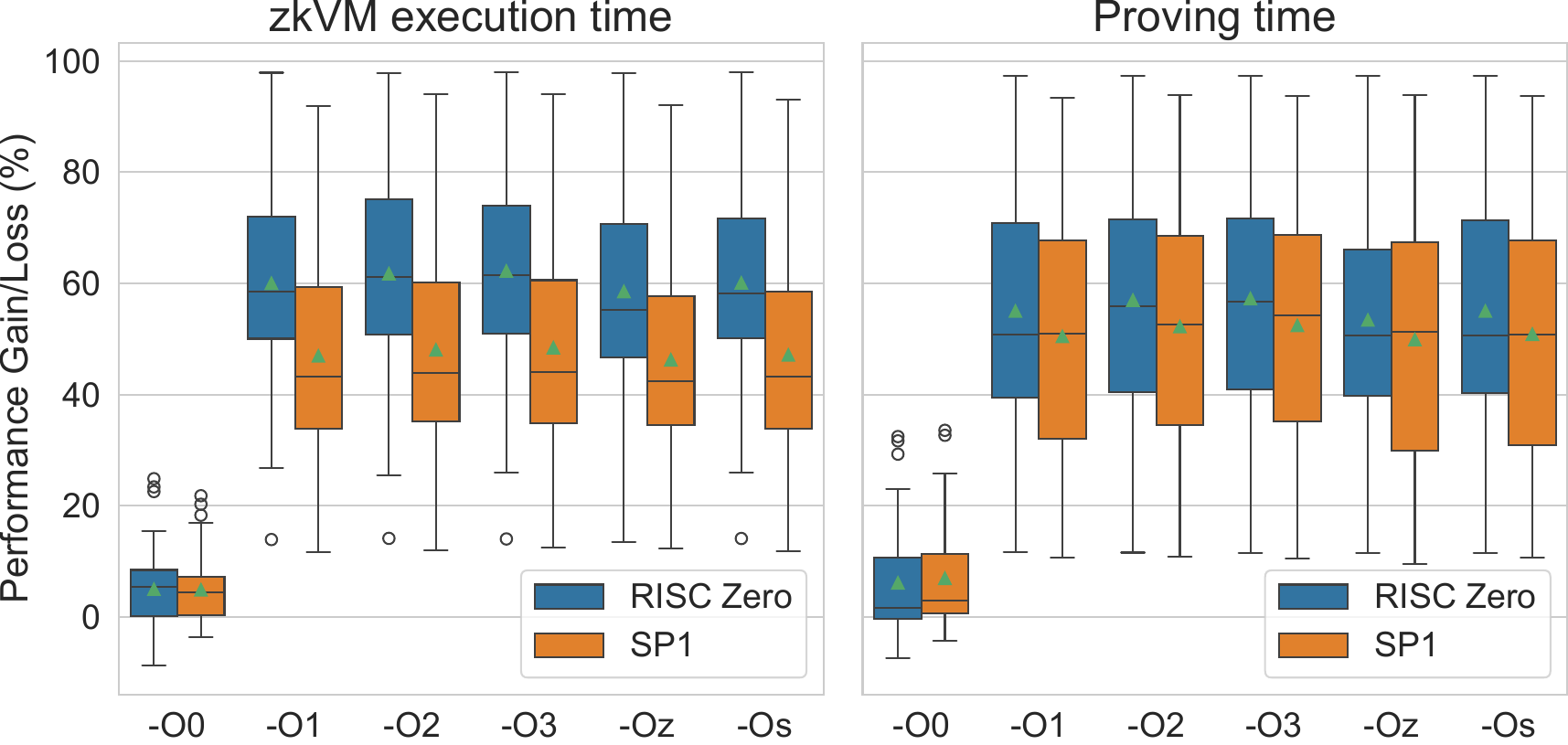}
\caption{Impact of standard optimization levels for zkVMs.}
\label{fig:standard-optimization-levels}
\end{figure}

\point{Standard optimization levels}
Figure~\ref{fig:standard-optimization-levels} shows
the impact of standard optimization levels
({\tt -Ox} flags)
on execution and proving time across both zkVMs,
relative to our baseline
where no optimizations are applied,
including Rust’s MIR optimizations
(Section~\ref{sec:research-questions}).
For brevity,
the effects on cycle count are omitted.
Excluding {\tt -O0},
no {\tt -Ox} level negatively impacts performance relative to the baseline.
In contrast,
{\tt -O0} (i.e., Rust's MIR optimizations)
causes regressions in~\empirical{19} programs on RISC Zero
and~\empirical{9} on SP1.
On average,
standard optimization levels improve execution time by
\empirical{60.5\%} on RISC Zero
and~\empirical{47.3\%} on SP1,
and proving time by~\empirical{55.5\%} and~\empirical{51.1\%},
respectively.
The {\tt -O3} level consistently achieves
the highest average performance gains,
while {\tt -Oz} (reducing binary size)
shows the lowest.



\point{Autotuning}
Overall,
the existing {\tt -Ox} flags
have a positive effect on both zkVM execution and proving time.
However, can we identify sequence of passes that
outperform them?
To explore this,
we employ OpenTuner~\cite{opentuner},
a genetic-based tuning framework
that finds an optimal combination of compilation flags
guided by a specific fitness function.
Although zkVM execution and proving time
are the ultimate metrics of interest,
they are too expensive to evaluate repeatedly
during autotuning.
Instead,
we use cycle count as a proxy,
as it is a natural performance predictor.
This choice offers two key advantages:
(1) cycle count avoids measurement noise,
such as the inherent imprecision of wall-clock timing
and (2) it is fast to compute
($\sim$1 second on average).

We autotune each benchmark independently
for each zkVM.
We consider all LLVM passes as tunable parameters,
creating sequences of passes up to a depth of~\empirical{20}.
For compiler options 
that receive integer or enum values
(e.g., {\tt -inline-threshold}),
we define appropriate tuning range
and use OpenTuner’s parameter types to
select the optimal value.
For most flags,
these ranges are large enough to be effectively unbounded.
Our artifact
~\footnote{\url{https://github.com/thomasgassmann/zkvm-compiler-optimizations}}
includes and documents
the tuning ranges for all examined compiler flags.
Each OpenTuner run performs~\empirical{160} iterations.
Even with this limited number of iterations,
the result of OpenTuner in several benchmarks outperforms {\tt -O3}.
On RISC Zero,
\empirical{18} out of~\tprograms programs outperform {\tt -O3}
in terms of cycle count reduction
(on average by~\empirical{6.3\%}),
while on RISC Zero,
\empirical{20} programs outperform {\tt -O3}
(on average~\empirical{14.6\%} cycle count reduction).

We further study 
the most common sub-sequences of passes
appearing in the best
and worst configurations
across all benchmarks and zkVMs.
Some findings align with the results of RQ1
(Section~\ref{sec:individual-passes}).
For example,
the {\tt inline} pass appears in
\empirical{573} out of~\empirical{580}
best-performing sequences,
where~\empirical{580} is the
total number of top-5 sequences considered
($2\ \text{zkVMs}\ \times 58\ \text{programs} \times 5\ \text{sequences each}$).
In contrast,
the {\tt licm} pass is present in~\empirical{385}
of the worst-performing ones.
This indicates
that {\tt inline} tends to be beneficial,
and {\tt licm} often detrimental,
even when combined with other passes.

\begin{figure}[t]
\centering
\begin{subfigure}{\linewidth}
\includegraphics[width=\linewidth]{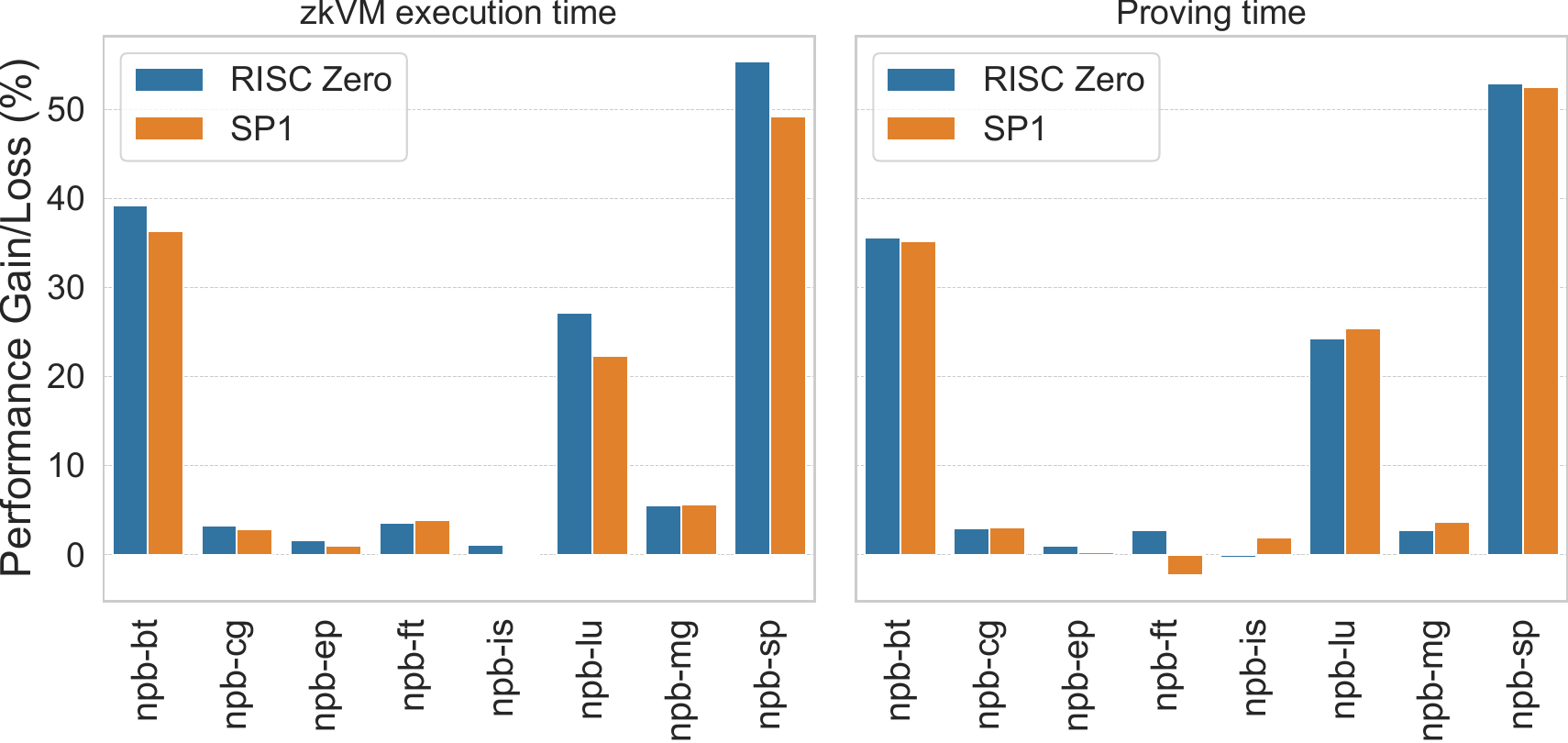}
\caption{NPB benchmark suite}
\vspace{3mm}
\label{fig:npb}
\end{subfigure}
\begin{subfigure}{\linewidth}
\includegraphics[width=\linewidth]{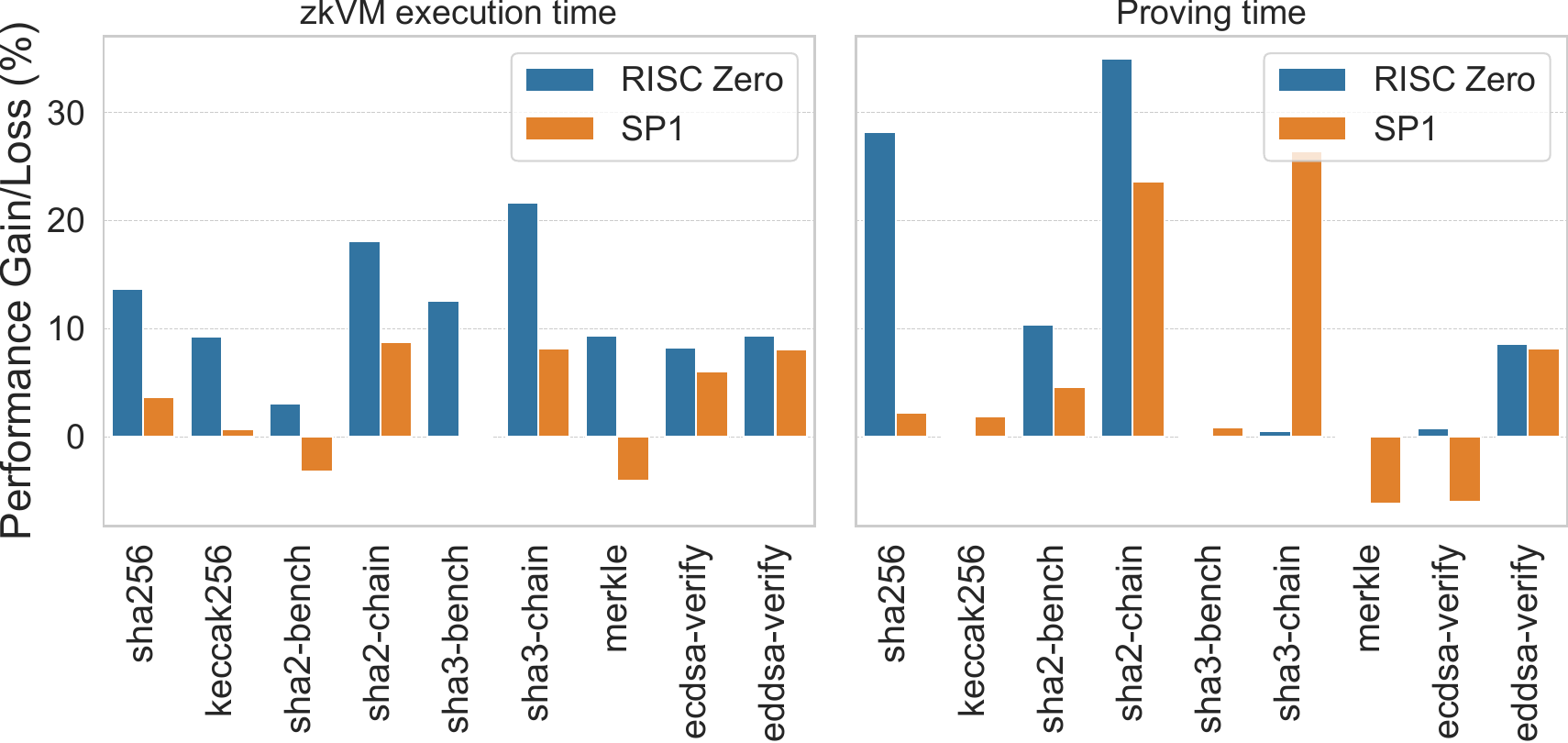}
\caption{Crypto benchmark suite}
\label{fig:crypto}
\end{subfigure}
\caption{
The speedup over {\tt -O3} (with LTO)
achieved by tuning optimization passes for each program
in the NPB and Crypto benchmark suites across both zkVMs.
}
\label{fig:autotune-improvement}
\end{figure}

However,
we also observed
more interesting behaviors:
{\tt inline} appears in~\empirical{122}
of the worst-performing sequences,
often when followed by {\tt licm}
(\empirical{23} out of~\empirical{122}).
In contrast,
the combination where {\tt licm} follows
{\tt inline} occurs~\empirical{50} times in
best-performing sequences.
Further analysis in Section~\ref{sec:analysis-selected-passes} explores
the root causes behind these context-sensitive effects.

\point{Autotuning specific benchmark suites}
In the previous experiment,
each benchmark was tuned individually
using cycle count as the fitness function.
Yet,
how does this translate
to zkVM execution and proving time?
To answer this,
we autotune (1) the NPB benchmark~\cite{npb-rust} suite
and (2) cryptography benchmarks
on both zkVMs,
running OpenTuner for a longer time frame
(\nnum{1600} iterations instead of~\empirical{160}).
We select the NPB suite for its large programs,
which are more likely to benefit from optimization,
and cryptography benchmarks
as they best represent typical zkVM workloads.

Figure~\ref{fig:autotune-improvement} shows
the performance improvement of each program
in the selected benchmark suites
relative to {\tt -O3} with link-time optimizations
(LTO).
For the NPB benchmark suite
(Figure~\ref{fig:npb}),
the autotuned configurations achieve
an average performance increase by~\empirical{17\%}
on SP1 in terms of both zkVM execution time and proving time.
On RISC Zero,
autotuning improves by~\empirical{19\%}
the execution time,
and by~\empirical{17\%} the proving time.
The {\tt npb-sp} program stands out,
as the result of the OpenTuner
achieves over $2\times$ speedup on both zkVMs for both metrics.

Moving to Figure~\ref{fig:crypto},
autotuning yields smaller,
yet meaningful,
gains for the cryptography benchmarks.
On average,
execution improves by~\empirical{11.6\%} on RISC Zero
and~\empirical{10.4\%} on SP1,
while proving time improves by~\empirical{3.5\%}
and~\empirical{6.8\%},
respectively. 
The smaller gains are expected,
as both zkVMs are already heavily optimized for
cryptographic workloads
through the use of~\emph{precompiles}. 
These are specialized,
built-in circuits that
implement certain expensive operations
(e.g., elliptic-curve signature verification or hashing)
directly in hardware-equivalent logic,
instead of executing
them step-by-step as regular instructions.
This allows these operations to be proven far more efficiently,
often in constant time.

Nevertheless,
even for benchmarks that make use of such precompiles,
such as {\tt ecdsa-verify},
{\tt eddsa-verify},
{\tt and keccak256},
autotuning can still deliver performance improvements
that exceed~\empirical{10\%}.

The findings confirm the viability of autotuning on zkVMs,
which has practical implications for
optimizing performance-critical programs
on zkVMs (see Section~\ref{sec:discussion}).

\point{Security-critical bug in SP1}
As a by-product,
our autotuning efforts uncover a
security-critical bug in SP1.
In particular,
while autotuning a specific program
from the Crypto benchmark suite,
we run into a certain sequence of passes
that leads to~\empirical{59\%} reduction in cycle count
compared to {\tt -O3} (w/ LTO)!
This,
however,
turns out to be a bug in SP1
that causes the zkVM to~\emph{silently} abort execution in mid-run
while still producing a proof that passes verification.
Such silent failures represent a serious vulnerability,
as they allow incorrect program results to be proven as correct.
We privately report this silent-failure
case to the SP1 development team,
who confirm it as a bug
and then patch it.

\subsection{RQ3: Comparison to Traditional CPUs}
\label{sec:x86-comparison}

\begin{figure}[t]
\centering
\includegraphics[width=0.9\linewidth]{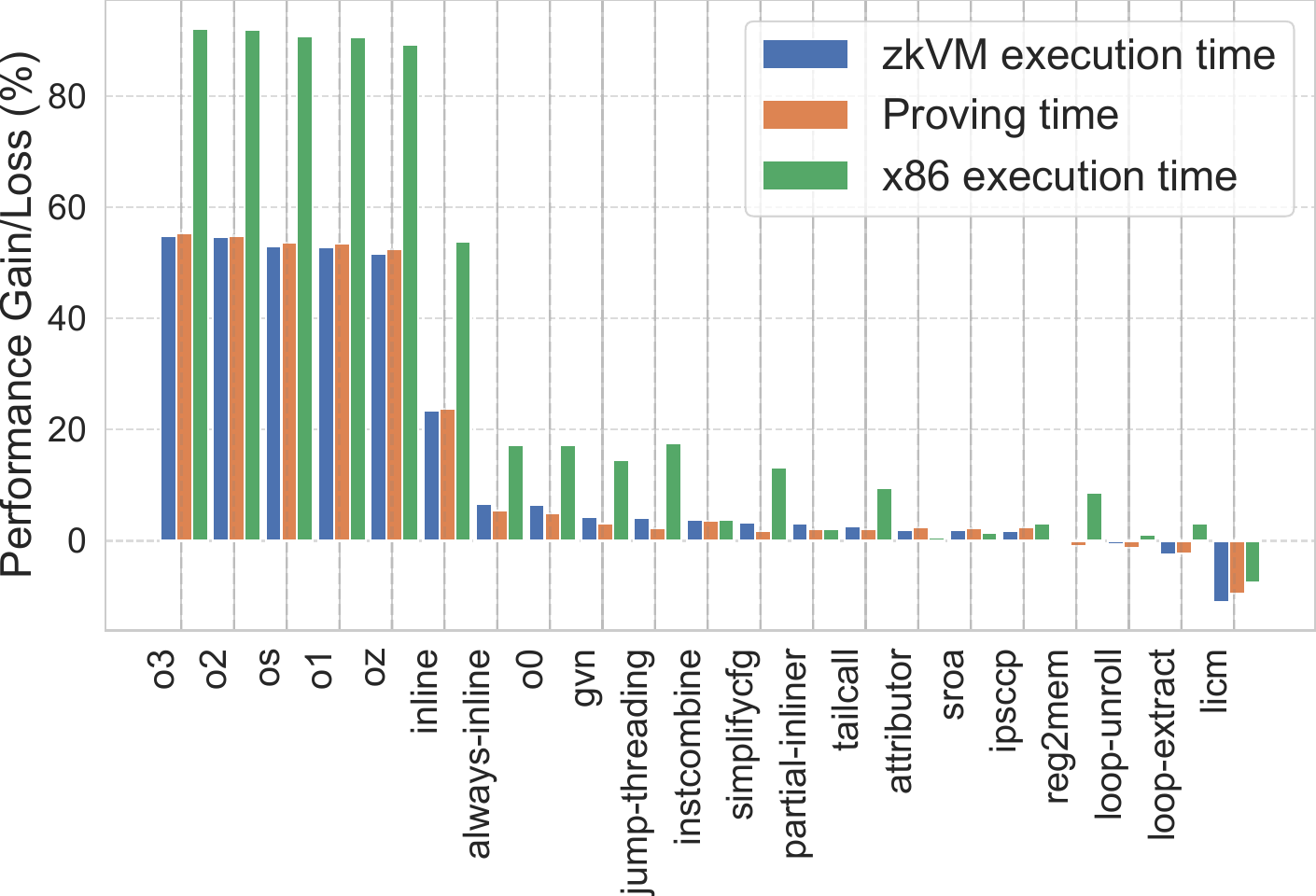}
\caption{
The average impact of each optimization on zkVM
and x86 performance,
measured across all benchmarks.
Passes with an average effect below 2\%
are omitted.
}
\label{fig:x86-comparison}
\end{figure}

We compare the impact of our studied LLVM passes
and flags on zkVMs versus a traditional architecture,
that is,
x86.

Figure~\ref{fig:x86-comparison} shows
the average impact of each pass on zkVM
and x86 performance,
measured across all benchmarks,
relative to the baseline
where no optimizations are applied,
including Rust's MIR optimizations.
For brevity,
we exclude passes with an average impact on zkVM execution time,
proving time,
and x86 execution time that is below~\empirical{2\%}.
For the majority of passes,
the direction of their impact (positive or negative)
is similar on both zkVMs and x86.
For example,
{\tt inline} is the most beneficial pass on both platforms,
while {\tt licm} is the most detrimental pass
when applied in isolation.
The latter occurs because {\tt licm} relies on
subsequent passes to eliminate
redundant {\tt getelementptr} operations,
which cause significant performance regressions
(Section~\ref{sec:analysis-selected-passes}).

However,
the magnitude of the effect is significantly larger on x86.
This could be explained by the fact
that many LLVM optimizations fail to show
their full potential on zkVMs.
This is mainly due to a mismatch in the cost model
or the underlying heuristics of LLVM,
which rely on assumptions
that hold only on the traditional architectures.
We provide further evidence on this
in Section~\ref{sec:analysis-selected-passes}.

\begin{figure}[t]
\centering
\includegraphics[width=0.8\linewidth]{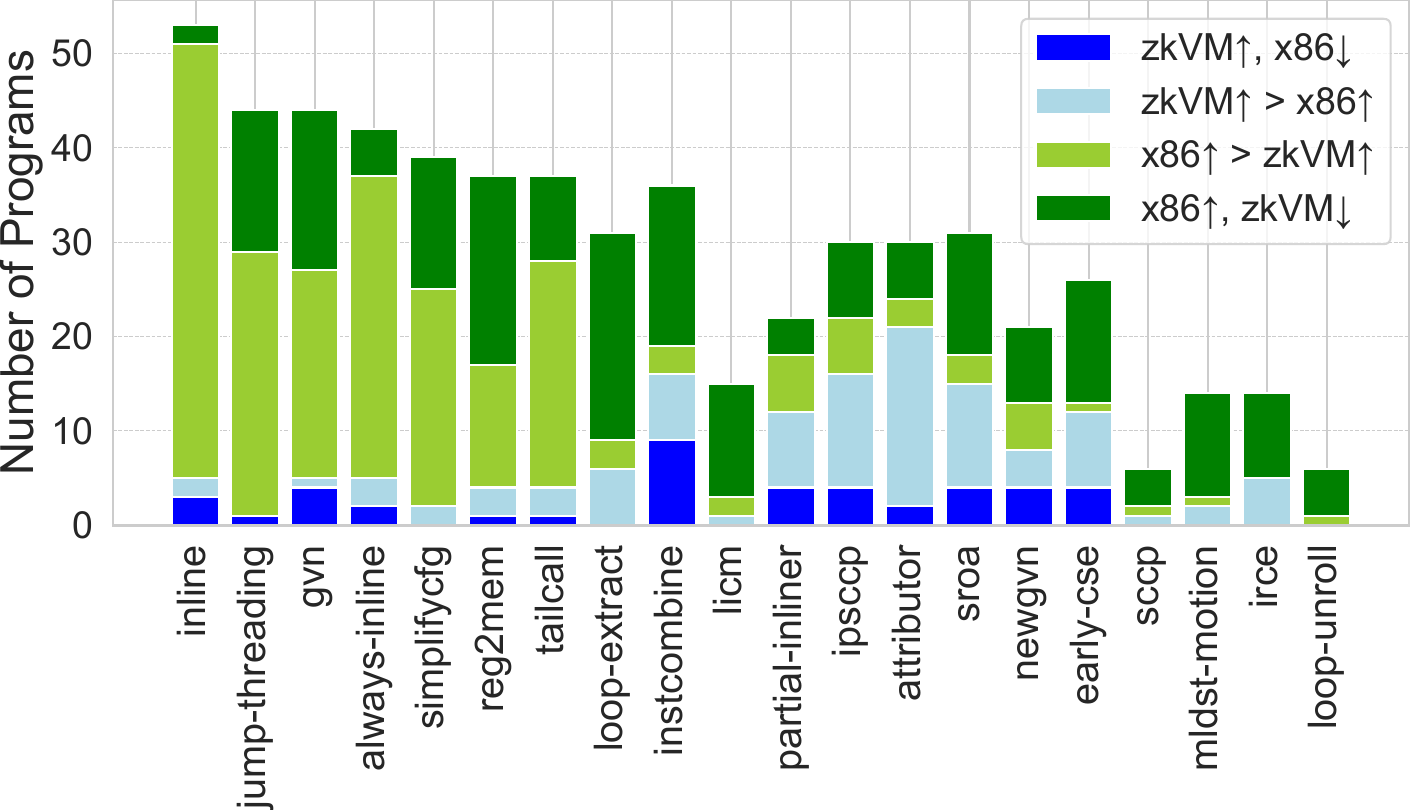}
\caption{
Number of programs where each pass exhibits:
(1) a performance gain on x86
but a loss on RISC Zero,
(2) a gain on both x86 and RISC Zero with >5\% higher gain on x86,
(3) a gain on both with >5\% higher gain on RISC Zero,
or (4) a gain on RISC Zero but a loss on x86.
}
\label{fig:x86-effects}
\end{figure}

%
%
%
%
%

Figure~\ref{fig:x86-effects} illustrates passes
with divergent effects on x86 vs. RISC Zero execution.
While most passes benefit both platforms,
the performance improvement is generally more pronounced on x86.
This exemplified by passes such as {\tt inline},
{\tt simplifycfg},
and {\tt jump-threading}.
Some passes,
such as {\tt reg2mem} and {\tt loop-extract},
tend to benefit x86 but degrade performance on RISC Zero.
In contrast,
passes like {\tt ipsccp} and {\tt attributor} show
stronger positive effects on RISC Zero.
The corresponding plots for SP1 and proving time follow
similar trends and are omitted for brevity.

\section{Performance Impact Analysis}
\label{sec:analysis-selected-passes}

We examine the results of our study
(Section~\ref{sec:results}) to distill
the key factors
that influence zkVM performance
and explain why certain optimization passes perform well or poorly.

\begin{figure*}[t]
\centering
\includegraphics[width=0.85\linewidth]{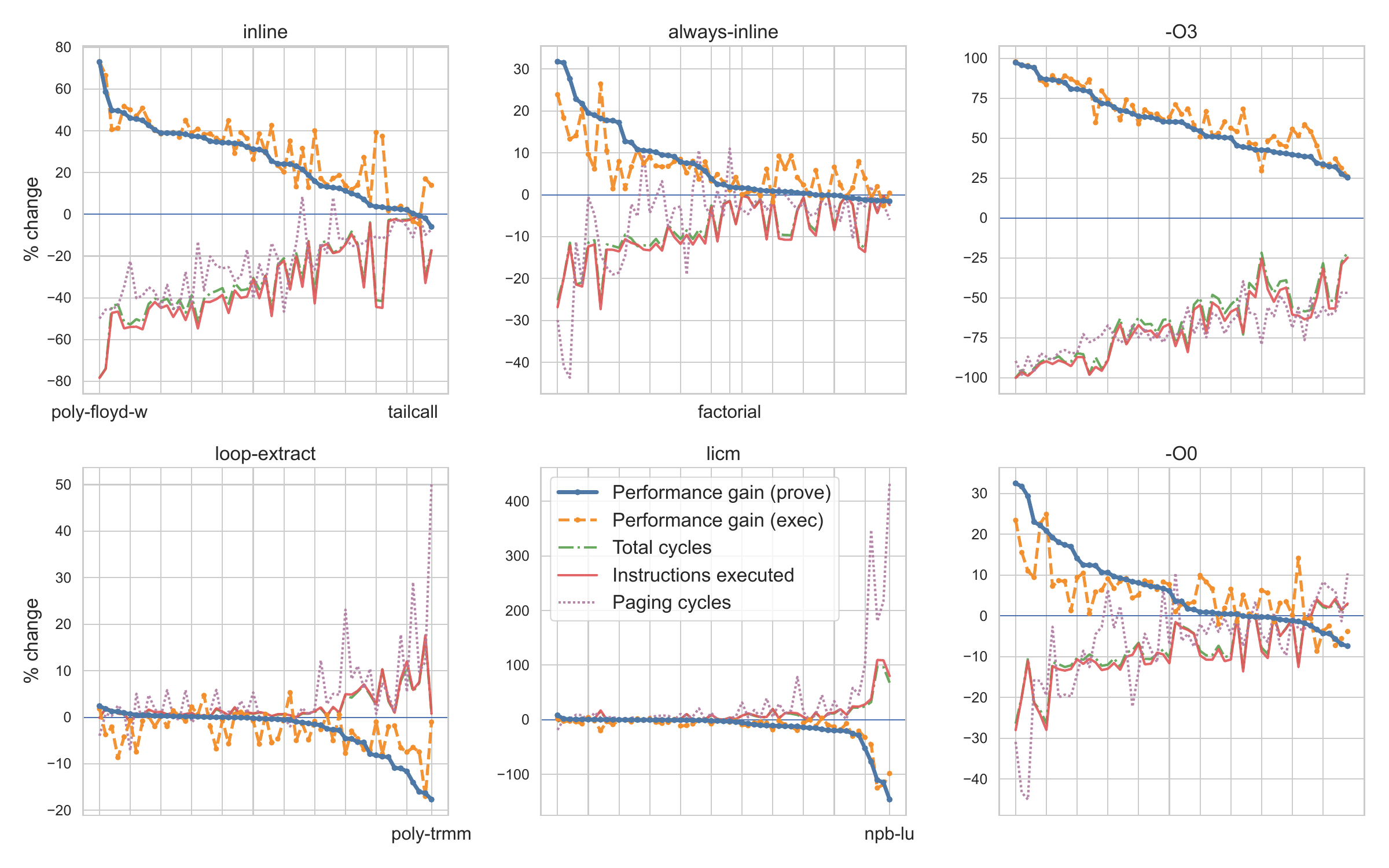}
\caption{
Impact of representative passes
on RISC Zero performance and cost metrics.
Each plot shows the percentage change relative to the baseline
for proving time and execution time,
along with changes in key zkVM cost components:
total cycles, dynamic instruction count,
and paging cycles.
The x-axis represents benchmarks.
For execution and proving time, higher values indicate
performance improvements,
whereas for the cost metrics,
higher values correspond to increased overhead.
}
\label{fig:root-cause-analysis}
\end{figure*}

\subsection{Cost Components}
\label{sec:cost-components}

According to RQ1
(Section~\ref{sec:individual-passes}),
{\tt inline} is the most beneficial pass for zkVMs,
while {\tt licm} is the most detrimental pass for
zkVM performance.
To understand~\emph{why},
we analyze the underlying cost components.
Beyond~\textit{cycle count},
which directly reflects execution cost
(Section~\ref{sec:individual-passes}),
we examine the~\textit{dynamic instruction count},
i.e., the number of executed instructions.
Our insight is that since zkVM's model
assigns almost a uniform cost for each instruction,
executing fewer instructions could potentially lead to
performance gains.
Finally,
for RISC Zero
(the corresponding metric is not available in SP1),
we consider~\textit{paging cycles},
which measure the cycles
required to move data between the guest program's
memory and the zkVM's storage system.
Page-ins and page-outs introduce
high overhead in zkVMs.
For example,
on RISC Zero a paging operation costs
roughly~\nnum{1130} cycles~\cite{risc0-guest-optimization-guide}.

Figure~\ref{fig:root-cause-analysis} illustrates
the impact of the two most beneficial passes
({\tt inline} and {\tt always-inline}),
and the two most detrimental passes
({\tt loop-extract} and {\tt licm})
on execution time
and proving time in RISC Zero,
relative to the baseline with no optimizations.
For completeness,
we also show the corresponding plots for {\tt -O3} and {\tt -O0}.
The other passes and pipelines exhibit
similar trends and can be
found in our artifact.
~\footnote{\url{https://github.com/thomasgassmann/zkvm-compiler-optimizations}}
The plots also report the effect of these passes on
key zkVM cost components:
cycle count,
dynamic instruction count,
and paging cycles.
There is a clear trend:
when a pass
(e.g., {\tt inline}) yields significant improvements in
execution or proving time,
it also leads to substantial reductions in the underlying cost metrics
(e.g., {\tt inline} on {\tt polybench-floyd-warshall}).
Similarly,
performance regressions are mainly attributed to
tremendous increases in either
dynamic instruction count or paging cycles,
or all of them
(e.g., {\tt licm} on {\tt npb-lu}).
When performance gains in execution or proving time are more modest,
the corresponding reductions in cost metrics are also limited,
or different metrics exhibit opposing effects,
such as a decrease in executed instructions accompanied
by an increase in paging cycles
(e.g., {\tt always-inline} on {\tt factorial}).

\begin{table}[t]
\caption{
Monotonic and linear relationships
between zkVM cost metrics and performance,
reported as average Kendall’s
and Pearson’s coefficients.
Correlations are computed per benchmark
over optimization variants.
}
\label{tab:correlations}
\resizebox{\linewidth}{!}{
\begin{tabular}{l|l|rrrr}
\toprule
                                     &                           & \multicolumn{2}{c}{\textbf{SP1}}        & \multicolumn{2}{c}{\textbf{RISC Zero}}  \\
\textbf{Performance metric}          & \textbf{Cost Metric}      & \textbf{Kendall's $\tau$} & \textbf{Pearson} & \textbf{Kendall's $\tau$} & \textbf{Pearson} \\
\midrule
\multirow{3}{*}{zkVM exec time} & Executed instr. & 0.87                 & 0.99             & 0.58                 & 0.97             \\
                                     & Paging cycles             & N/A                  & N/A              & 0.38                 & 0.88             \\
                                                                          & Total cycles              & 0.87                 & 0.99             & 0.59                 & 0.97             \\
\midrule
\multirow{3}{*}{Proving time}        & Executed instr. & 0.49                 & 0.9              & 0.5                  & 0.95             \\
                                     & Paging cycles             & N/A                  & N/A              & 0.33                 & 0.89             \\
                                                                          & Total cycles              & 0.49                 & 0.89             & 0.48                 & 0.95             \\
\midrule
Paging cycles                        & Executed instr. & N/A                  & N/A              & 0.43                 & 0.9            \\ 
\bottomrule
\end{tabular}}
\end{table}

\point{Monotonicity and linearity}
To assess whether
these cost components can reliably guide
compiler optimization decisions for zkVMs,
we evaluate the monotonic relationship
between dynamic instruction count,
paging cycles,
and zkVM performance.
To this end,
for each benchmark,
we compute Kendall’s $\tau$ correlation coefficient~\cite{kendall}
between each cost metric
(dynamic instruction count,
paging cycles, total cycles)
and both zkVM execution time and proving time,
across individual passes applied in isolation.
We also report the Pearson correlation coefficient,
which captures linear relationships.
Table~\ref{tab:correlations} summarizes the results.

Overall,
dynamic instruction count
and paging cycles exhibit moderate-to-strong positive correlations
with both execution time
and proving time across zkVMs.
For example,
a Kendall’s $\tau$ value of~\empirical{0.58}
between dynamic instruction count 
and RISC Zero execution time indicates that,
for a given benchmark,
there is a~\empirical{79\%} probability
that an optimization
increasing dynamic instruction count
also increases execution time.
This probability is computed by $(1 + \tau) / 2$,
where $\tau$ is the Kendall's coefficient.
The high Pearson coefficients ($>$ 0.9) further suggests
that this relationship is linear.

For RISC Zero,
dynamic instruction count shows
stronger correlation with both execution time
and proving time than paging cycles.
While dynamic instruction count is the primary cost driver,
the two metrics are complementary,
as reflected by a Kendall’s $\tau$ of 0.43 between them.
In~\empirical{28.5\%} of cases,
the metrics disagree:
an optimization may reduce instruction count
while increasing paging overhead,
or vice versa
(e.g., Figure~\ref{fig:root-cause-analysis};
{\tt always-inline} on {\tt factorial}).

\begin{figure}[t]
\centering
\begin{subfigure}{0.45\linewidth}
\begin{verbatim}
const N: usize = ...;
let mut v = [0; N];
for k in 0..N {
  v[k] = 42;
}
\end{verbatim}
\caption{A simple loop of depth one.}
\label{fig:licm-nesting1}
\end{subfigure}
\hspace{1.5mm}
\begin{subfigure}{0.51\linewidth}
\begin{verbatim}
const N: usize = ...;
let mut v = ...;
for k in 0..N {
 for j in 0..N {
  for i in 0..N {
   for l in 0..N {
    v[k][j][i][l] = 42;
... }
\end{verbatim}
\caption{Nested loops of depth four.}
\label{fig:licm-nesting4}
\end{subfigure}
\caption{Simplified examples
that cause significant performance
regressions when using {\tt licm}.}
\label{fig:licm-examples}
\end{figure}

\subsection{Case Studies}
\label{sec:case-studies}

Based on our study's results
(Section~\ref{sec:results}),
we examine a selection of passes
with interesting performance behaviors.
We select these passes based
on four criteria:
(1) passes that cause considerable performance losses on zkVMs,
(2) passes that provide significant performance gains on zkVMs,
(3) passes that benefit x86 execution
but degrade performance on zkVMs,
and (4) passes that improve both x86
and zkVM performance,
but with a substantially greater impact on x86.
Our analysis aims to identify
the root causes of these performance behaviors
based on the main zkVM cost components
(Section~\ref{sec:cost-components}).
Then,
we derive a set of principles that
can guide potential LLVM improvements
for better zkVM performance
(Section~\ref{sec:zkvm-specific-optimizations}).
\label{sec:case-studies}

\point{Significant performance regressions on zkVMs}
{\tt licm} is the most detrimental pass for zkVM performance in RQ1
(Section~\ref{sec:individual-passes}).
The effect is particularly severe on
the {\tt npb-lu} benchmark,
where {\tt licm} slows down proving time by
$\empirical{2.7}\times$ on RISC Zero
and~\empirical{$1.7\times$} on SP1,
and execution by~\empirical{$1.7\times$} on both.
The root cause is a sharp increase in~\emph{paging} cycles:
+\empirical{444\%} on RISC Zero,
+\empirical{69\%} on SP1.
Indeed,
across all benchmarks,
{\tt licm} increases paging cycles by~\empirical{32\%}.

Figure~\ref{fig:licm-nesting1} illustrates
a simplified loop extracted
from {\tt npb-lu} where {\tt licm} incurs this overhead.
This comes from
the increased number of
{\tt getelementptr} instructions in the LLVM IR,
each computing the address of an array element
to store the constant {\tt 42} (line 4).
These extra memory operations
in turn cause significant paging pressure with
an increase in page-ins and page-outs.

\begin{figure}[t]
\centering
\begin{verbatim}
pub fn work(x: u64) -> u64 {
  let mut sum = x;
  for j in 0..100 {
    sum = sum.wrapping_mul(31).wrapping_add(j);
  }
  sum
}
fn main() {
  let n = 1000;
  for i in 0..n { let res = work(i as u64); }      
}
\end{verbatim}
\caption{Example Rust function that {\tt inline}
causes a performance regression for zkVMs.}
\label{fig:inline-example}
\end{figure}

The pattern is also exemplified
by the program in Figure~\ref{fig:inline-example}
(extracted from the {\tt tailcall} program).
Here,
inlining causes major regressions
on both zkVM execution
($0.8\times$ speedup)
and proving
($0.45\times$ speedup).
The root cause is stack spilling:
RISC-V has only 32-bit registers,
so each 64-bit variable occupies two.
After inlining the function {\tt work}
(lines 1 and 10),
three {\tt u64} variables
(outer-loop index {\tt i} on line 10,
inner-loop index {\tt j} on line 3,
and {\tt sum} on line 2) must coexist.
This forces the compiler to spill both {\tt j} and {\tt sum}
to the stack every inner-loop iteration.
This doubles the {\tt lw}/{\tt sw} (load/store) instructions,
which account for most of the $2\times$ cycle increase.
Indeed,
the number of the executed instructions increases by~\empirical{102\%},
while the number of paging cycles increases by~\empirical{54\%}.

\textit{Principle 1 (P1):
Optimizations that introduce a lot of paging pressure should be avoided},
as it harms zkVMs.
Examples of passes that can potentially increase paging pressure
include {\tt licm},
{\tt inline},
and {\tt reg2mem}.
Since page-ins and page-outs are highly expensive in zkVMs,
compilers targeting zkVMs must prioritize paging-aware
cost models when transforming loops or spilling to memory.

To validate this principle,
we manually increase
the nesting of the loop of Figure~\ref{fig:licm-nesting1}.
Nested loops exacerbate the negative effects
of paging pressure:
with a nesting of depth 4
(Figure~\ref{fig:licm-nesting4}),
{\tt licm} increases paging cycles and dynamic instruction count by
\empirical{46\%} and~\empirical{155\%},
respectively,
compared to \empirical{7\%} and~\empirical{25\%},
at a depth 2.
The extra memory pressure stemming from {\tt licm}
is evident 
when loops iterate over arrays,
as this leads to a higher number of
load/store operations.

\point{Significant improvements on zkVMs}
According to RQ1 (Section~\ref{sec:individual-passes}),
{\tt inline} is the most beneficial pass for zkVMs.
But is LLVM’s default inlining optimal~\cite{optimal-inlining}?
To find out, we raise the inlining threshold
from 225 (default) to 4328
using the {\tt -inline-threshold} option,
allowing the compiler to inline
more aggressively even larger functions.
Notably,
the value of 4328 comes from
the OpenTuner’s autotuning results
on the NPB and crypto benchmarks
(Section~\ref{sec:combinations-optimizations}).
Compared to {\tt -O3} with the default threshold,
the higher {\tt -inline-threshold} improves
execution by~\empirical{6\%} on RISC Zero
and~\empirical{1\%} on SP1 on average,
with some benchmarks showing dramatic gains,
such as +\empirical{44\%} on RISC Zero and +\empirical{40\%} on SP1 for
the {\tt npb-bt} program.
These performance gains are accompanied by
similar decreases in both dynamic instruction count and
paging cycles.
In contrast,
increasing the inline threshold degrades
x86 performance on average by~\empirical{1\%}.

\begin{figure}[t]
\centering
\begin{verbatim}
pub fn matmul(mat: &[[f64; 5]; 5], vec: &[f64; 5]) {
  let mut res = [0.0; 5];
  for col in 0..5 {
    for row in 0..5 {
      res[row] += mat[col][row] * vec[col];
... }
\end{verbatim}
\caption{Example Rust function that computes the matrix-vector product of a $5 \times 5$ matrix and a $5$-dimensional vector.}
\label{fig:loop-unroll-example-rust}
\end{figure}

\textit{Principle 2 (P2):
Applying inlining selectively based on dynamic instruction count
is beneficial for zkVMs.}
Our analysis in Figure~\ref{fig:root-cause-analysis} shows
that reductions in executed instructions
caused by {\tt inline} often lead to
substantial zkVM improvements,
primarily by eliminating call overhead.
However,
as shown in Figure~\ref{fig:inline-example},
inlining should be avoided
if it causes spills to the stack,
as load/store instructions 
increase dynamic instruction count
and may trigger expensive memory paging
(Figure~\ref{fig:inline-example}).
This suggests that compilers targeting zkVMs should:
(1) adopt a zkVM-aware cost model for inlining rather than
reusing heuristics from traditional CPUs,
and (2) tune inlining thresholds per workload,
especially for performance-critical code
(Section~\ref{sec:discussion}).

To validate this principle,
we manually inline selected functions
in the {\tt tailcall} benchmark by
adding {\tt \#[inline(always)]} annotations,
while excluding the function that triggers stack spills.
This reduces dynamic instruction count by~\empirical{5\%},
yielding a~\empirical{2\%} faster execution
and eliminating the previous regression.

\point{Improvements on x86; degradations on zkVMs}
On {\tt polybench-durbin},
the {\tt loop-unroll} pass improves
x86 execution by~\empirical{9\%}
but degrades RISC Zero performance by~\empirical{8.7\%}.
Figure~\ref{fig:loop-unroll-example-rust} shows
an oversimplified version of this benchmark
where {\tt loop-unroll} produces this divergence
between x86 and zkVM performance.
Overall,
x86 sees gains of 9\% when loop unrolling this example program
compared to the unoptimized baseline,
whereas both zkVMs consistently
suffer slowdowns of 3\%--10\%
in both execution and proving time.
The regressions are caused by
a~\empirical{2\%} increase in executed instructions
and~\empirical{17\%} increase in paging cycles
on both zkVMs.

\begin{table}[t]
\centering
\caption{
Relative speedup and change in dynamic instruction count for
4$\times$ and 16$\times$ loop unrolling,
compared to the non-unrolled assembly code.
}
\label{tab:loop-unroll-asm-speedups}
\resizebox{\linewidth}{!}{
\begin{tabular}{l|r|rrrrrr}
\toprule
\multirow{3}{*}{\textbf{Unrolling Factor}}
& \multirow{1}{*}{\textbf{x86}} & \multicolumn{3}{c}{\textbf{SP1}} & \multicolumn{3}{c}{\textbf{RISC Zero}} \bigstrut \\
& & \textbf{\#instr.} & \textbf{prove} & \textbf{exec} & \textbf{\#instr} & \textbf{prove} & \textbf{exec} \bigstrut \\
\midrule
    4 & +28.1\% & -50\% & +24.3\% & +6.6\% & -50\% & +51.4\% & +39.5\% \\
    16 & +31.5\% & -63\% & +26.4\% & +1.5\% & -62\%  & +51.6\% & +52.7\% \\
\bottomrule
\end{tabular}}
\end{table}

When used with {\tt -O3} and a high {\tt unroll-threshold}
(aggressive unrolling),
loop unrolling becomes beneficial
compared with {\tt -O3} and no loop unrolling.
Specifically,
on RISC Zero,
execution improves by~\empirical{16\%}
and proving by~\empirical{3.7\%};
on SP1,
by~\empirical{0.5\%} and~\empirical{1.7\%}.
In both zkVMs,
these performance gains are attributed to
by~\empirical{4--5\%} decrease in cycle count
(explained by a 5\% decrease in dynamic instruction count).
However,
these gains are still modest compared to
x86’s $32\times$ speedup improvement.
This trend also appears in {\tt polybench-gemm},
which is also a matrix multiplication benchmark.
By increasing the unrolling threshold,
we save around~\empirical{43--44\%} in terms of cycles
(compared to {\tt -O3}),
by executing \empirical{44\%} fewer instructions.
This leads to a~\empirical{11.5\%}
improvement in execution time on RISC Zero
and \empirical{2\%} improvement on SP1.

\textit{Principle 3 (P3):
Performance gains are observed only
when loop unrolling leads to a reduction in the executed instructions.}
Since zkVMs cannot exploit instruction-level parallelism,
out-of-order,
or superscalar execution,
performance gains when unrolling loops
depend entirely on reducing instruction count.
Compilers targeting zkVMs should
prioritize instruction reduction
over traditional hardware assumptions.

To validate this principle,
we manually unroll loops in RISC-V assembly
of program shown in Figure~\ref{fig:loop-unroll-example-rust},
avoiding interference from other passes
or vectorization.
Table~\ref{tab:loop-unroll-asm-speedups} reports
speedups for $4\times$ and $16\times$ loop unrolls.
This manual effort yields substantial performance gains
for both zkVMs,
with RISC Zero even surpassing the gains seen on x86.
These improvements come mainly
from executing fewer loop bookkeeping instructions,
such as counter increments,
comparisons,
and branch jumps.
As dynamic instruction count decreases,
paging cycles also decrease,
by about~\empirical{40\%} on average.

\begin{figure}[t]
\centering
\begin{subfigure}[t]{\linewidth}
\begin{verbatim}
fn abs_i32_branchy(x: i32) -> i32 {
  if x < 0 { -x } else { x }
}
\end{verbatim}
\caption{Rust function computing $|x|$ using a branch.}
\end{subfigure}

\begin{subfigure}[b]{0.48\linewidth}
\begin{verbatim}
bltz a0, .LBB52_2
ret
.LBB52_2:
neg a0, a0
ret
\end{verbatim}
\caption{RISC-V before applying \texttt{simplifycfg}.}
\label{fig:simplifycfg-before}
\end{subfigure}
\hfill
\begin{subfigure}[b]{0.48\linewidth}
\begin{verbatim}
srai a1, a0, 31
xor a0, a0, a1
sub a0, a0, a1
ret
\end{verbatim}
\caption{RISC-V after applying \texttt{simplifycfg}.}
\label{fig:simplifycfg-after}
\end{subfigure}
\caption{An example Rust function
including its assembly output
before and after applying \texttt{simplifycfg}.}
\label{fig:simplifycfg-rust}
\end{figure}


\point{Substantial improvement on x86; minimal effect on zkVMs}
On {\tt polybench-nussinov},
{\tt simplifycfg} improves x86 by~\empirical{23.6\%},
but SP1 execution by only~\empirical{2.6\%}.
Figure~\ref{fig:simplifycfg-rust} shows
a minimized function from this benchmark
that computes the absolute value of an integer using a branch.
{\tt simplifycfg} replaces the branch with equivalent arithmetic operations
(Figures~\ref{fig:simplifycfg-before} and~\ref{fig:simplifycfg-after}).
When tested with an array of integers chosen at random,
the optimized version runs $2.2\times$ faster on x86,
presumably thanks to less branch misprediction overhead.
However,
on zkVMs,
the optimized code increases cycle count
by~\empirical{17.7\%} on RISC Zero
and~\empirical{7.6\%} on SP1.
These are primarily explained
by increases in
the number of executed instructions (\empirical{8\%})
and paging cycles (\empirical{17\%}).
As a result,
proving time degrades by
\empirical{18.1\%} and~\empirical{7\%}
on RISC Zero and SP1,
respectively,
while execution time worsens by~\empirical{8.5\%} and~\empirical{7.7\%}.


\textit{Principle 4 (P4):
Branches should not always be eliminated,
as branch elimination may increase
the number of executed instructions
by requiring both paths to be evaluated
}.
Compilers targeting zkVMs should be more conservative
when removing branches,
for two reasons.
First,
branches are relatively cheap
and incur no misprediction penalty
compared to traditional CPUs.
Second,
predicated execution increases proving cost,
since both paths must be proven and executed.

To validate this principle,
we modify the unoptimized binary of
{\tt polybench-nussinov}
by replacing the implementation of the {\tt max} function
with a branchless version
implemented with right-shifts and bitwise {\tt AND}
operations.
While this change yields a~\empirical{4\%}
performance improvement on x86,
it degrades performance on RISC Zero by~\empirical{4.4\%}
and on SP1 by~\empirical{0.5\%}.
This regression is caused by
a~\empirical{3\%} increase in dynamic instruction count
on both zkVMs.

\section{Implications}
\label{sec:implications}

We demonstrate the practical value of our study's results
by refining existing LLVM passes to
improve zkVM performance
and then extract several key lessons from our findings.

\subsection{zkVM-Specific Optimizations}
\label{sec:zkvm-specific-optimizations}

Existing LLVM optimization levels already boost
zkVM performance (Section~\ref{sec:combinations-optimizations}),
but our root cause analysis
(Section~\ref{sec:analysis-selected-passes}) reveals
significant untapped potential.
Building on these insights,
we propose a~\emph{non-exhaustive} set of lightweight backend
and pass modifications
across three dimensions to
further unlock LLVM’s potential for zkVMs.

\point{Change set 1: Adopting a zkVM-specific cost model}
The default RISC-V cost model in LLVM is not zkVM-aware,
yet instruction costs on zkVMs differ
greatly from traditional CPUs.
For example,
multiplication costs the same as addition,
and division is not expensive.
Memory access is also far more expensive
when it triggers page-ins or page-outs
(see~\textit{Principle 1}, Section~\ref{sec:analysis-selected-passes}).
We address these by updating the
{\tt RISCVTTIImpl} and {\tt RISCVTargetLowering} classes,
which provide target-specific cost models to LLVM’s middle-end
so that passes can make
better decisions for this architecture.

\point{Change set 2: Updating defaults and heuristics}
Many LLVM passes employ heuristics
that are tuned for traditional architectures.
This tuning may not be suitable for zkVMs.
Adjusting these for zkVM-specific instruction costs,
such as applying function inlining
(\textit{Principle 2})
or loop unrolling
(\textit{Principle 3})
only when it reduces instructions count
can improve the performance of the generated code.

We increase the default inlining threshold
and adjust parameters
such as {\tt inline-call-penalty} accordingly.
We also update {\tt simplifycfg} to make
branch elimination more conservative
and avoiding evaluation of both sides of a predicated operation
(\textit{Principle 4}).

\point{Change set 3: Disabling irrelevant passes}
Our RQ1 analysis
(Section~\ref{sec:individual-passes}) shows
that some passes offer little
or no benefit on zkVMs,
mainly because they rely on hardware features
that are unsupported in zkVMs.
For example,
{\tt speculative-execution} hoists side-effect-free instructions
to reduce branch-misprediction latency
on out-of-order,
superscalar CPUs.
This advantage is irrelevant to zkVMs,
which execute RISC-V code strictly
in order inside an arithmetic circuit.
Similarly,
we also disable {\tt loop-data-prefetch},
and {\tt hot-cold-splitting},
as they provide no measurable gain.

All these targeted LLVM changes took~\emph{only}
a few hours to implement
and required fewer than 100 lines of code.

\begin{figure}[t]
\centering
\begin{subfigure}{\linewidth}
\centering
\includegraphics[width=0.9\linewidth]{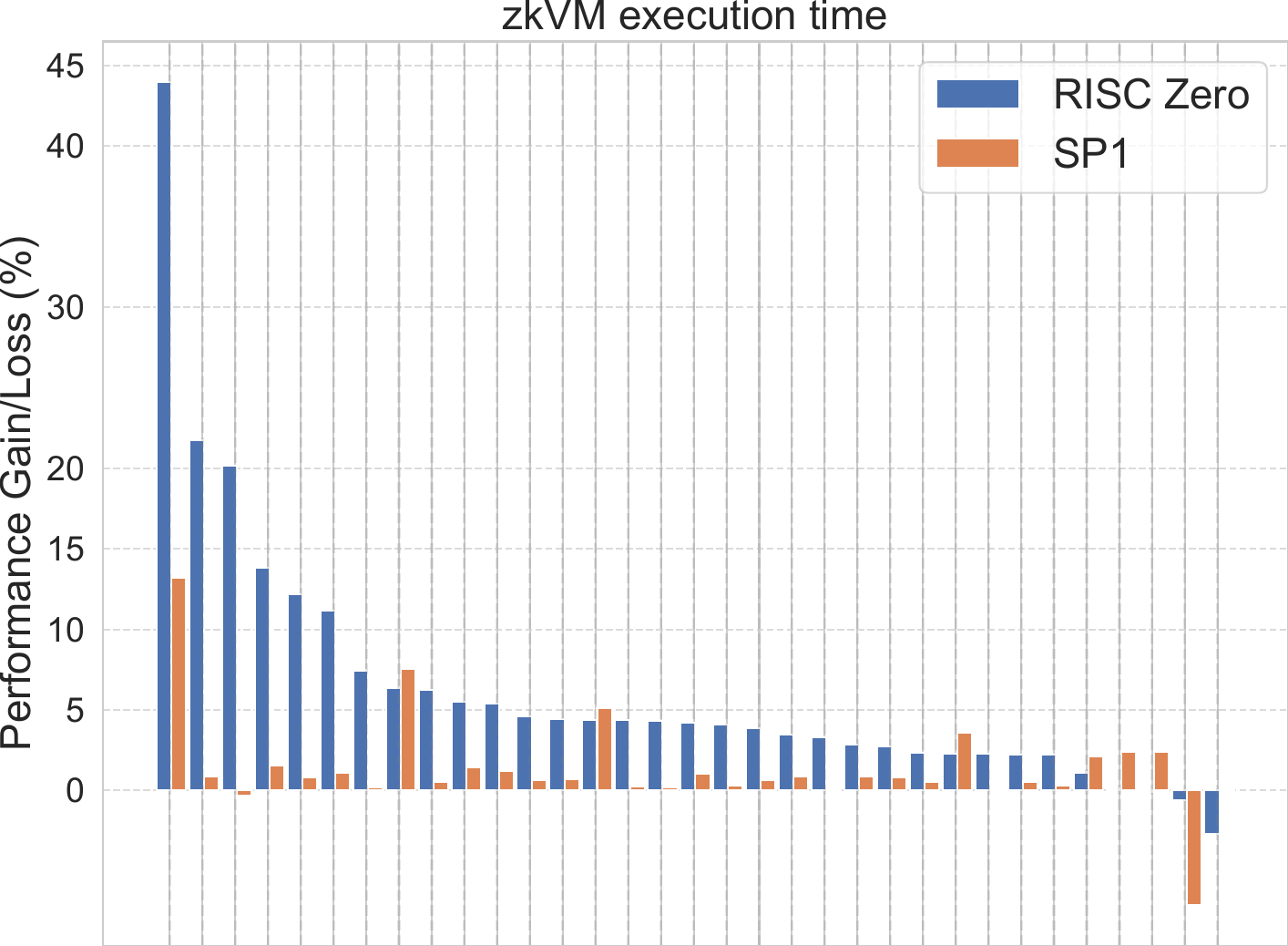}
\end{subfigure}
\begin{subfigure}{\linewidth}
\centering
\includegraphics[width=0.9\linewidth]{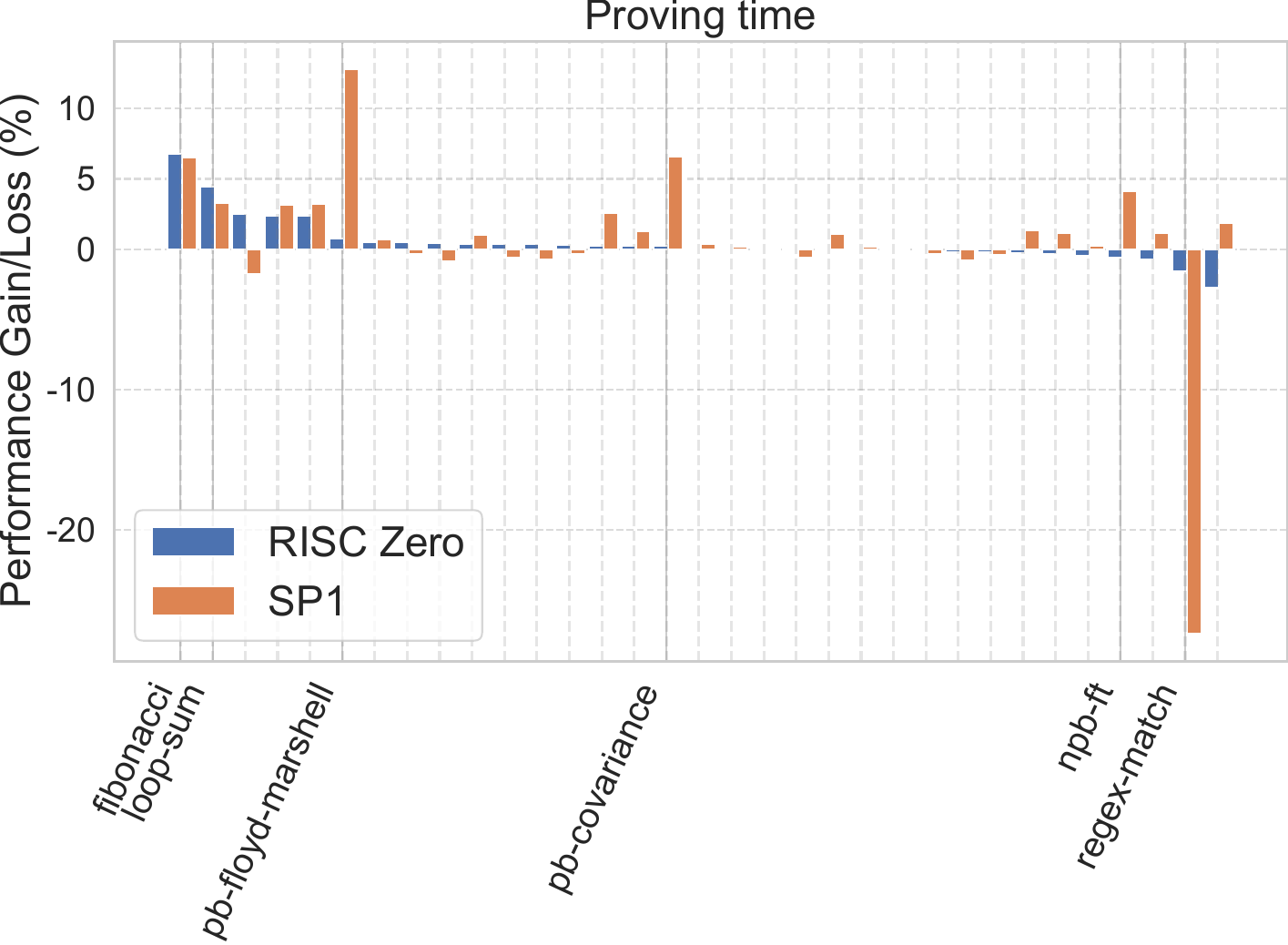}
\end{subfigure}
\caption{
    Changes in zkVM execution performance
    {\tt -O3} with our modified LLVM compared to
    {\tt -O3} with the original LLVM.
    Programs with effects below 2\% are omitted.
}
\label{fig:llvm-modifications-prove}
\end{figure}

\point{Results}
We evaluate {\tt -O3} in our modified LLVM
against {\tt -O3} in the original toolchain.
Figure~\ref{fig:llvm-modifications-prove} shows
that our LLVM changes improve zkVM performance
across a wide range of benchmarks,
with execution-time gains exceeding~\empirical{10\%} in many cases
and reaching up to~\empirical{45\%} on RISC Zero for {\tt fibonacci}.
This improvement directly validates
our cost model and core principle
(Section~\ref{sec:cost-components}):
reducing dynamic instruction count translates to
proportional performance gains.
The {\tt fibonacci} benchmark experiences
a~\empirical{50\%} reduction in dynamic instruction count,
which corresponds to
the~\empirical{45\%} execution time improvement on RISC Zero.
Specifically,
our division cost-model update enables
LLVM to select a~\emph{single} {\tt remu} instruction instead of
a~\emph{series} of bitwise operations for remainder calculation.

The modified LLVM outperforms vanilla {\tt -O3} on
execution time in~\empirical{39}/\empirical{58} benchmarks
on RISC Zero
(avg. +\empirical{4.6\%})
and~\empirical{19}/\empirical{58} on SP1
(avg.~\empirical{+1\%}),
with only two and one regressions,
respectively.
On SP1,
proving time improves for~\empirical{27}/\empirical{58} programs
by more than~\empirical{1\%},
with only three regressions.
On RISC Zero,
most proving-time effects are small,
with 8 programs improving by more than~\empirical{1\%}
and 4 programs showing slight regressions.
One outlier is {\tt regex-match} on SP1,
where our modifications lead to a~\empirical{27.3\%}
regression in proving time.
This is caused by our
changes in the {\tt inline}'s heuristics,
which make SP1 produce 20 proof shards instead of 16.
Proof sharding divides the program into smaller proving units;
more shards may increase proving time 
due to additional aggregation overhead.

Notably,
our performance improvements are
mainly explained by dynamic instruction count reduction:
our modifications reduce
the number of the executed instructions
by~\empirical{2.6\%} on average across both zkVMs,
with only three benchmarks showing slight increases
(max +\empirical{0.5\%}). 
\subsection{Lessons Learned and Discussion}
\label{sec:discussion}

\textit{\textbf{zkVM users can rely on the compiler to optimize their programs.}}
Our results show that zkVM users can already benefit
from LLVM’s existing optimization levels (e.g., {\tt -O3}),
which improve both execution
and proving time by over~\empirical{40\%}
(RQ2, Figure~\ref{fig:standard-optimization-levels}).
However\dots

\textit{\textbf{\dots the magnitude of these gains
is smaller compared to what similar optimizations
achieve on traditional architectures.}}
Our fine-grained pass-level analysis
(Section~\ref{sec:individual-passes}) reveals
that many passes still rely on
hardware-centric assumptions and heuristics
that harm zkVM performance
(e.g., branch elimination,
Section~\ref{sec:analysis-selected-passes}).
By refining just a few of these passes to be zkVM-aware,
we already surpass LLVM’s default {\tt -O3}
(Figure~\ref{fig:llvm-modifications-prove}).
This proof-of-concept highlights that even small,
yet targeted zkVM-specific compiler changes can deliver
substantial speedups
and motivate the development of
a dedicated zkVM-aware LLVM backend.

Section~\ref{sec:analysis-selected-passes} covers only a few passes
(e.g., {\tt licm}, {\tt inline}).
Many others could unlock further optimization opportunities.
For example,
{\tt instcombine} improves half the benchmarks
but hurts the rest
(Figure~\ref{fig:prog-impact}).
Performing a similar root-cause analysis on this pass could reveal
which rewrites of {\tt instcombine} to keep and which to disable.

\textit{\textbf{Autotuning performance-critical software.}}
As part of RQ2 (Section~\ref{sec:combinations-optimizations}),
we also show the viability of program autotuning on zkVMs.
This has direct,
tangible value in real-world settings.
For example,
in the Ethproofs ecosystem~\cite{ethproofs},
where different zkVMs race
to prove Ethereum blocks in real-time 
(i.e., in less than 12-seconds -- Ethereum's block time).
Despite significant progress, 
real-time proving still requires substantial GPUs, 
and much research and engineering work 
focuses on achieving even minor improvements.
Exploring autotuning for prover workloads
on real Ethereum blocks is
thus a promising and economically impactful direction.

Another takeaway from our autotuning experiment
is that it can yield significant gains
even for programs using precompiles,
with improvements of up to~\empirical{10\%}.
This suggests that compiler-level tuning can complement zkVMs’
built-in accelerations by
further reducing proving costs.

\textbf{\textit{Profile-guided optimization.}}
High-quality profiling data is critical
for generating efficient code.
For example,
static profile-guided optimization could significantly
improve inlining decisions
and reduce instruction count.
Profiling data could also guide the design of
future zkVM-specific passes.
For example,
a page coalescing pass could place hot variables
within as few 1 KB pages as possible,
minimizing expensive page-ins and page-outs on zkVMs,
which we identify in Section~\ref{sec:analysis-selected-passes}
as a major bottleneck.

\textit{\textbf{Combined cost metrics.}}
Our analysis in Section~\ref{sec:cost-components}
establishes that zkVM performance is primarily
driven by dynamic instruction count.
Paging cycles provide complementary information.
This motivates the use of combined cost models
that integrate multiple components into
a single performance predictor.
While this two-factor model explains the majority of
performance variation (Table~\ref{tab:correlations}),
some effects remain unexplained,
such as SP1's proof sharding overhead
in Section~\ref{sec:zkvm-specific-optimizations}.
Future work could refine the model by
incorporating more zkVM-specific factors
(e.g., use of {\tt precompiles}),
or learning workload-specific weights
(e.g., memory-heavy vs. computation-heavy programs).

\textit{\textbf{Use of superoptimization.}}
Since zkVM execution and proving are
orders of magnitude more expensive
than native execution,
it is often worth trading longer compilation time for faster proving.
Superoptimization is a technique
that allows compilers to do so
by finding the optimal sequence of instructions
for implementing a given computation.
Superoptimizers that operate on LLVM IR
(e.g., Souper~\cite{souper})
could be adapted to zkVMs.
Incorporating a zkVM-specific cost model
would make them a powerful tool
for generating highly efficient zkVM guest code.



\section{Related Work}
\label{sec:related}
\point{zkVMs}
\citet{SNARK-vonNeumann} 
introduce the first system 
to prove executions of a von Neumann machine, 
albeit on a \emph{bounded} machine 
with a predetermined cycle count. 
The field has since moved to \emph{unbounded} proving, 
e.g., Cairo~\cite{Goldberg2021CairoA} which leverages STARKs~\cite{STARK} 
to scale. 
Recent efforts improve zkVM performance 
via better circuit generation 
and prover back-ends 
as well as more efficient proof systems~\cite{jolt,lasso,Goldberg2021CairoA,sp1,risc0}. 
Our work complements these improvements 
by studying how compiler optimizations interact with zkVM cost models, 
and outline directions 
to reduce proving overhead 
through traditional (or novel) optimizations. 

\point{Benchmarking ZKPs}
Benchmarking ZKPs is crucial for identifying bottlenecks 
and guiding improvements. 
\citet{zkbench} introduce \textsc{zk-bench}, 
a framework for benchmarking ZK DSLs~\cite{circom}. 
\citet{benchzkevms} 
provide the first large-scale systematic study of ZK-EVMs, 
highlighting design trade-offs and bottlenecks. 
Industry efforts benchmark zkVMs 
to compare capabilities 
and performance~\cite{a16z-benchmarks,succinct-zkvm-perf,aligned-zkvm-bench}. 
We build on these benchmarks 
by reusing selected industry workloads 
and complement them 
with a broad and diverse Rust/C suite. 


\point{Autotuning \& Superoptimization}
\citet{bintuner} 
develops BinTuner to
investigate the impact of compiler optimizations 
on binary code difference. 
BinTuner is based on OpenTuner~\cite{opentuner}, 
a framework for building domain-specific autotuners
using genetic algorithms.
In this work, 
we leverage OpenTuner 
to show that autotuning is a viable method to improve the 
proving performance of zkVMs for a given program.
A complementary line of work is 
\emph{superoptimization}~\cite{stoke,souper,superoptimization}, 
which searches for optimal instruction sequences 
for a given (typically loop-free) fragment. 
Exploring superoptimization for zkVMs is
a promising future work,
and our study provides valuable insights to guide it
(Section~\ref{sec:discussion}).

\section{Conclusion}
\label{sec:conclusion}
We presented 
the first systematic study 
of how compiler optimizations 
affect two production RISC-V zkVMs.
Most passes help, 
yet gains are modest 
and a few are harmful (e.g., \texttt{licm}). 
Autotuning yields up to $2.2\times$ proving speedups 
beyond \texttt{-O3}.
We established a strong correlation
between dynamic instruction count
and zkVM performance and,
based on this insight,
prototype zkVM-aware LLVM changes
that further improve zkVM performance.

Future work can extend this study to additional zkVMs 
and proof systems, 
unifying cost metrics, 
and exploring superoptimizations for zkVMs. 
These compiler-centric directions can push zkVM proving performance 
beyond what proof-system research and engineering alone can deliver.

\bibliographystyle{ACM-Reference-Format}
\bibliography{main}

\appendix                

\section{Differences between zkVMs and Native Execution}
\label{app:zkvm-vs-cpu}

The following considers two popular zkVMs (RISC Zero~\cite{risc0} and SP1~\cite{sp1}) and summarizes the key differences compared to traditional architectures.

\point{Instruction latency} On traditional architectures, instruction latency can vary significantly depending on the instruction type, cache hits/misses, and other micro-architectural effects \cite{instruction-tables-traditional-architectures}. In zkVMs, however, most instructions have a uniform cost \cite{risc0-guest-optimization-guide}, with most instructions taking exactly the same number of cycles.     

\point{Memory access} Memory access is relatively cheaper on zkVMs. For example, on RISC Zero, memory access costs only one cycle if the page is already paged in. Page-ins and page-outs on RISC Zero are, however, more expensive, taking around 1130 cycles on average \cite{risc0-guest-optimization-guide}. zkVMs also do not have any caches making memory access always synchronous.

\point{Control flow} zkVMs do not have branch prediction or instruction caches, which means that control flow instructions can be cheaper on zkVMs, as there is no penalty for mispredicted branches.

\point{Out-of-order execution} zkVMs do not have out-of-order execution. They also do not have super-scalar execution, instruction-level parallelism, or pipelining, which means that all instructions are executed sequentially.

\point{Floating-point operations} zkVMs do not support native floating-point operations, which means that all floating-point operations must be emulated. This can lead to significant overhead, especially for complex floating-point operations. 

\point{Multithreading} zkVM execution is always single-threaded. There is no parallelism in the execution.

\point{Precompiles} zkVMs provide precompiles for expensive operations such as cryptographic hashes or elliptic-curve arithmetic. Precompiles can significantly reduce the cost of these operations compared to implementing them in the guest program.

\begin{figure}[t]
    \centering
    \includegraphics[width=0.5\textwidth]{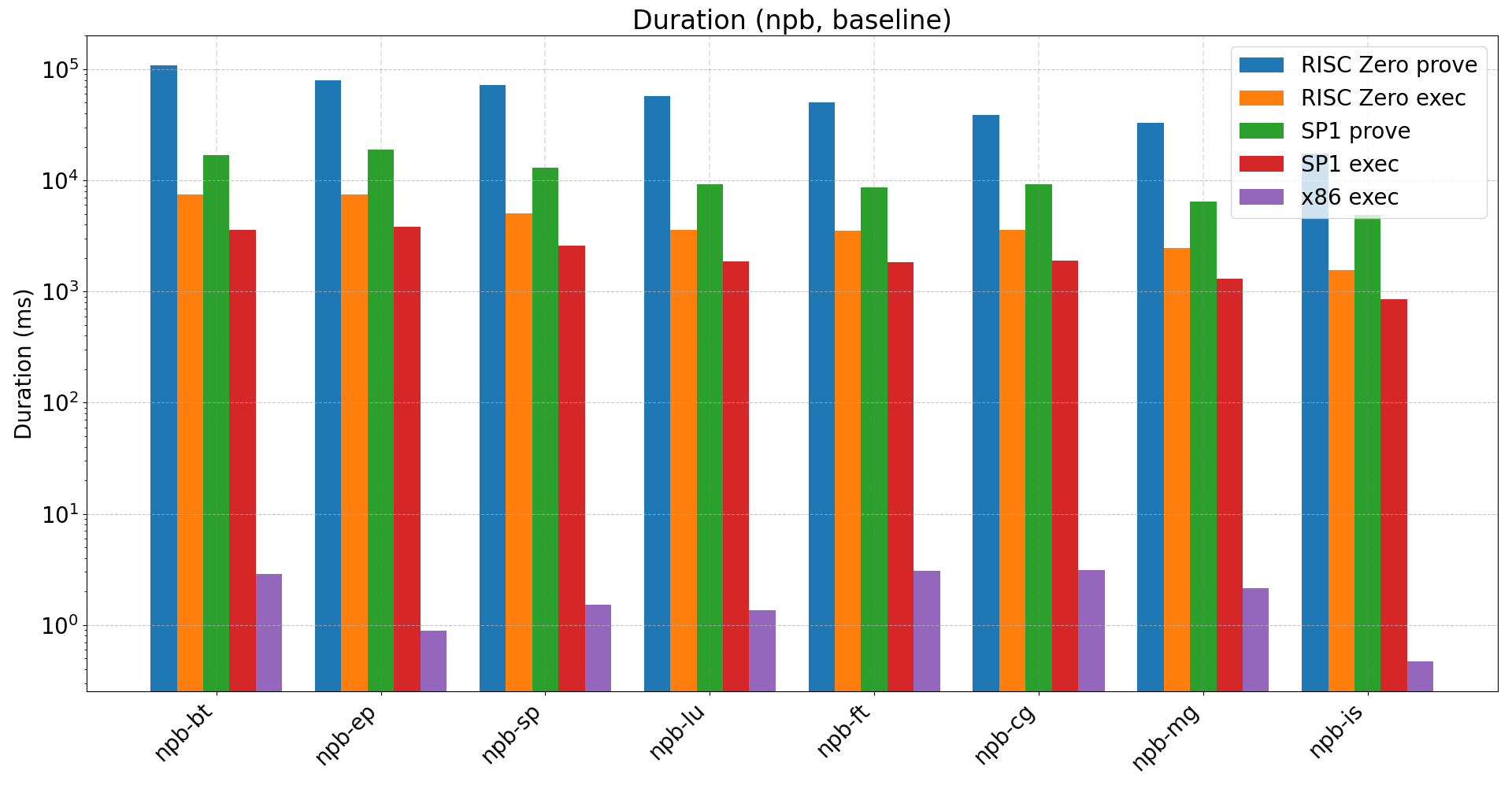}
    \caption{Median duration of zkVM operations compared to
    execution time on x86 for the NPB benchmark suite.
    All binaries were compiled with no optimizations.}
    \label{fig:zkvm-costs}
\end{figure}

zkVM execution and proving are order of magnitude slower
than native execution.
Figure~\ref{fig:zkvm-costs} illustrates this
by considering the time spent on zkVM execution,
proof generation,
and native execution for every program
included in the NPB benchmark suite~\cite{npb-rust}.

\begin{table*}[!t]
\centering
\caption{Benchmark programs and their SLOC counts, split across two subtables with balanced rows.}
\captionsetup[subtable]{justification=centering}
\setlength{\tabcolsep}{3pt}
\begin{subtable}[t]{0.44\textwidth}
\centering
\footnotesize
\begin{tabular}{llccc}
\toprule
\textbf{Program} & \textbf{Libraries used} & \textbf{Precomp.} & \textbf{Rust} & \textbf{Total} \\
\midrule
bigmem & None & No & 32 & 32 \\
ecdsa-verify & k256 & Yes & 16 & 16 \\
eddsa-verify & ed25519\_dalek & Yes & 16 & 16 \\
factorial & None & No & 26 & 26 \\
keccak256 & sha3 & Yes & 13 & 13 \\
loop-sum & None & No & 29 & 29 \\
merkle & rs\_merkle & No & 46 & 46 \\
npb-bt & None & No & 2915 & 2915 \\
npb-cg & None & No & 827 & 827 \\
npb-ep & None & No & 485 & 485 \\
npb-ft & None & No & 1151 & 1151 \\
npb-is & None & No & 815 & 815 \\
npb-lu & None & No & 2746 & 2746 \\
npb-mg & None & No & 2121 & 2121 \\
npb-sp & None & No & 2728 & 2728 \\
polybench-2mm & None & No & 113 & 113 \\
polybench-3mm & None & No & 148 & 148 \\
polybench-adi & None & No & 107 & 107 \\
polybench-atax & None & No & 78 & 78 \\
polybench-bicg & None & No & 80 & 80 \\
polybench-cholesky & None & No & 63 & 63 \\
polybench-correlation & None & No & 98 & 98 \\
polybench-covariance & None & No & 84 & 84 \\
polybench-deriche & None & No & 146 & 146 \\
polybench-doitgen & None & No & 83 & 83 \\
polybench-durbin & None & No & 67 & 67 \\
polybench-fdtd-2d & None & No & 93 & 93 \\
polybench-floyd-warshall & None & No & 59 & 59 \\
polybench-gemm & None & No & 90 & 90 \\
polybench-gemver & None & No & 116 & 116 \\
& & & & \\
\bottomrule
\end{tabular}
\end{subtable}\hfill
\begin{subtable}[t]{0.53\textwidth}
\centering
\footnotesize
\begin{tabular}{llcccc}
\toprule
\textbf{Program} & \textbf{Libraries used} & \textbf{Precomp.} & \textbf{Rust} & \textbf{C/C++} & \textbf{Total} \\
\midrule
polybench-gesummv & None & No & 81 & 81 \\
polybench-gramschmidt & None & No & 88 & 0 & 88 \\
polybench-heat-3d & None & No & 82 & 0 & 82 \\
polybench-jacobi-1d & None & No & 64 & 0 & 64 \\
polybench-jacobi-2d & None & No & 70 & 0 & 70 \\
polybench-lu & None & No & 64 & 0 & 64 \\
polybench-ludcmp & None & No & 103 & 0 & 103 \\
polybench-mvt & None & No & 79 & 0 & 79 \\
polybench-nussinov & None & No & 99 & 0 & 99 \\
polybench-seidel-2d & None & No & 70 & 0 & 70 \\
polybench-symm & None & No & 88 & 0 & 88 \\
polybench-syr2k & None & No & 85 & 0 & 85 \\
polybench-syrk & None & No & 81 & 0 & 81 \\
polybench-trisolv & None & No & 67 & 0 & 67 \\
polybench-trmm & None & No & 74 & 0 & 74 \\
regex-match & regex & No & 45 & 0 & 45 \\
rsp & serde, & Yes & 11 & 0 & 11 \\
& rsp-client-executor, & & & & \\
& c-kzg, & & & & \\
& bytemuck\_derive, & & & & \\
& bincode & & & & \\
sha2-bench & sha2 & No & 36 & 0 & 36 \\
sha2-chain & sha2 & No & 44 & 0 & 44 \\
sha3-bench & sha3 & No & 35 & 0 & 35 \\
sha3-chain & sha3 & No & 44 & 0 & 44 \\
sha256 & None & No & 163 & 0 & 163 \\
spec-605 & None & No & 48 & 2159 & 2207 \\
spec-619 & None & No & 53 & 203 & 256 \\
spec-631 & None & No & 46 & 6496 & 6542 \\
tailcall & None & No & 56 & 0 & 56 \\
zkvm-mnist & None & No & 181 & 0 & 181 \\
\bottomrule
\end{tabular}
\end{subtable}
\label{tab:sloc-counts}
\end{table*}

\section{Benchmark Suite Details}
\label{app:benchmarks}

In the following, we give a brief summary of the benchmark programs.

\point{PolyBench} PolyBench is a benchmark suite of 30 numerical computations from domains like linear algebra, image processing, physics simulations, dynamic programming, statistics, and more. Originally written in C \cite{polybench-c}, we use the Rust port \cite{polybench-rust} of the benchmark suite. The inputs of the benchmarks have been reduced to fit within the constraints of the zkVMs.

\point{NPB} The NAS Parallel Benchmarks (NPB) are a set of eight benchmarks designed to evaluate the performance of parallel supercomputers originally developed by NASA's Advanced Supercomputing Division \cite{npb-c}. Martins et al. \cite{npb-rust} port the benchmarks to Rust and provide a sequential version of the benchmarks. We use the sequential version of the Rust port for our evaluation. Similar to PolyBench, the input sizes have been reduced to fit within the constraints of the zkVMs.

\point{SPEC CPU 2017} The SPEC CPU benchmark suite \cite{spec-cpu} is a widely used benchmark suite for evaluating the performance of computer systems. Due to the complexity of the benchmarks and the fact that they are written in C/C++, whose support is still limited in zkVMs, we only use a subset of three benchmarks: \href{https://www.spec.org/cpu2017/Docs/benchmarks/605.mcf_s.html}{\texttt{605}}, \href{https://www.spec.org/cpu2017/Docs/benchmarks/619.lbm_s.html}{\texttt{619}} and \href{https://www.spec.org/cpu2017/Docs/benchmarks/631.deepsjeng_s.html}{\texttt{631}}. These three benchmarks were chosen based on their ease of integration.

\point{a16z crypto zkVM benchmarks} Andreessen Horowitz (a16z) crypto has published a set of benchmarks \cite{a16z-benchmarks} comparing the RISC Zero, SP1 and Jolt \cite{jolt} zkVMs. We use the benchmarks from their repository, which consist of a set of cryptographic operations (\texttt{sha2}, \texttt{sha3}) as well as an allocation-heavy benchmark (\texttt{bigmem}).

\point{Succinct Labs benchmarks} We also incorporate a subset of the zkVM benchmarks released by Succinct Labs \cite{succinct-zkvm-perf}. The benchmarks we include consist of a set of cryptographic operations (\texttt{ecdsa-verify}, \texttt{eddsa-verify}, \texttt{keccak256}) as well as a Fibonacci sequence benchmark (\texttt{fibonacci}).

\point{Reth Succinct Processor (RSP)} We also adopt the Reth Succinct Processor (RSP) \cite{succinct-reth} benchmarks from Aligned Inc.'s benchmark suite \cite{aligned-zkvm-bench}. RSP represents a common type of workload for zkVMs that generates zero-knowledge proofs for EVM (Ethereum Virtual Machine) block execution. We use block 20526624 as input for our benchmarks.\footnote{The FAQ on \href{https://github.com/succinctlabs/rsp/blob/249b34ee0c5307b59dc48f1f45a474b04669c6c4/README.md}{RSP's GitHub page} highlights block 20526624 as "a good small block to test". While adapting our benchmark setup to test a different block is relatively straightforward, running the benchmarks for all optimization profiles is time-consuming. The impact of compiler optimizations on RSP under consideration of different blocks is left as future work.}

\point{Others} In addition to the above benchmarks, we also add some additional benchmarks: a benchmark training a neural network on the MNIST dataset (\texttt{zkvm-mnist}) downsampled to 7x7 images, a benchmark performing regex matching (\texttt{regex-match}), a benchmark performing a merkle tree inclusion proof (\texttt{merkle}), another SHA-256 benchmark (\texttt{sha256}) as well as three smaller benchmarks (\texttt{factorial}, \texttt{loop-sum} and \texttt{tailcall}).

Overall, the 58 benchmark programs that we have assembled span a diverse range of workloads and input characteristics, ranging from numerical computations to memory-intensive operations, cryptographic primitives commonly used in zkVMs, and real-world applications such as RSP. This variety ensures a thorough evaluation of compiler optimizations across all major zkVM use cases.

\point{Benchmark Programs}
Table \ref{tab:sloc-counts} summarizes the benchmark programs used in this evaluation, including their source lines of code (SLOC) counts for Rust and C/C++. The SLOC counts of any libraries used by the benchmarks are \emph{not} included in the table.

A list of the library versions is listed in Table \ref{tab:library-versions}. All programs also use the crates \texttt{sp1-zkvm} (version 4.1.1) and \texttt{risc0-zkvm} (version 1.2.3).

\begin{table*}[th]
\centering
\caption{Programs and their associated libraries and versions.}
\label{tab:library-versions}
\captionsetup[subtable]{justification=centering}
\setlength{\tabcolsep}{4pt}
\begin{subtable}[t]{0.47\textwidth}
\centering
\footnotesize
\begin{tabular}{lll}
\toprule
\textbf{Program} & \textbf{Library} & \textbf{Version/Commit} \\
\midrule
PolyBench \cite{polybench-rust} & -- & \href{https://github.com/JRF63/polybench-rs/tree/babdd974bf92f03fbe4b2146ca356b4561f447d7}{babdd97} \\
NPB \cite{npb-rust} & -- & \href{https://github.com/glbessa/NPB-Rust/tree/4bd879ba4086fdf2a269af12699a01f500dc6c11}{4bd879b} \\
ecdsa-verify & k256 & 0.13.3 \\
eddsa-verify & ed25519\_dalek & 2.1.1 \\
merkle & rs\_merkle & 1.5.0 \\
regex-match & regex & 1.11.1 \\
sha2-bench & sha2 & 0.10.8 \\
sha2-chain & sha2 & 0.10.8 \\
\bottomrule
\end{tabular}
\end{subtable}\hfill
\begin{subtable}[t]{0.47\textwidth}
\centering
\footnotesize
\begin{tabular}{lll}
\toprule
\textbf{Program} & \textbf{Library} & \textbf{Version/Commit} \\
\midrule
sha3-bench & sha3 & 0.10.8 \\
sha3-chain & sha3 & 0.10.8 \\
\multirow{6}{*}{rsp} & serde & 1.0.204 \\
& rsp-client-executor & \href{https://github.com/succinctlabs/rsp/commits/249b34ee0c5307b59dc48f1f45a474b04669c6c4/}{249b34e} (SP1) \\
& rsp-client-executor & \href{https://github.com/succinctlabs/rsp/commits/4ceefdfb74c1691dd009168f0c7aec3d746ef8b3/}{4ceefdf} (RISC Zero) \\
& c-kzg & 1.0.3 \\
& bytemuck\_derive & 1.8.0 \\
& bincode & 1.3.3 \\
\bottomrule
\end{tabular}
\end{subtable}
\end{table*}

\section{Statistics of Baseline}
\label{app:baseline-stats}

\begin{table*}[t]
\centering
\caption{
Statistics of execution and proving times
(in seconds) for each zkVM across all benchmarks,
measured on the unoptimized baseline
with no compiler passes applied.
}
\label{tab:baselinestats}
\footnotesize
\begin{tabular}{lrrrrrrrr}
\toprule
\multirow{2}{*}{\textbf{zkVM}} & \multicolumn{4}{c}{\textbf{Execution}} & \multicolumn{4}{c}{\textbf{Proving}} \\
\cmidrule(lr){2-5} \cmidrule(lr){6-9}
 & \textbf{Min} & \textbf{Max} & \textbf{Mean} & \textbf{Median} & \textbf{Min} & \textbf{Max} & \textbf{Mean} & \textbf{Median} \\
\midrule
RISC Zero & 0.04 & 157.70 & 4.51 & 0.34 & 0.53 & 2071.24 & 60.85 & 3.83 \\
SP1       & 0.06 & 41.81  & 1.70 & 0.23 & 0.38 & 205.87  & 8.89  & 1.90 \\
\bottomrule
\end{tabular}
\end{table*}

Table~\ref{tab:baselinestats} shows
the statistics of execution and proving times
(in seconds) for each zkVM across all benchmarks,
measured on the unoptimized baseline
with no compiler passes applied.

\end{document}

%% file: packages.tex
\usepackage{graphicx}
\usepackage{subcaption}
\usepackage{booktabs}
\usepackage{caption}
\usepackage{colortbl}
\usepackage{xcolor}
\usepackage{multirow}
\usepackage{enumitem}
\PassOptionsToPackage{table,dvipsnames}{xcolor} 
\usepackage{sepnum}
\usepackage{xspace}
\usepackage{bigstrut}

\definecolor{kw}{rgb}{0,0,0.7}
\definecolor{hl}{gray}{0.88}

\usepackage{listings}
\usepackage{longtable}
\usepackage{booktabs}
\usepackage{multirow}
\usepackage{tikz}
\usepackage{adjustbox}
\usetikzlibrary{arrows.meta, positioning, shapes.multipart, fit, calc, backgrounds}

\usepackage{hyperref}
\hypersetup{
    colorlinks=true,
    linkcolor=blue,
    filecolor=magenta,
    urlcolor=blue,
    citecolor=blue
    }

%% file: macros.tex
\newcommand{\point}[1]{\noindent\textbf{#1.}}
\newcommand{\tpasses}{\nnum{64}\xspace}
\newcommand{\tprograms}{\nnum{58}\xspace}
\newcommand{\empirical}[1]{#1}
\newcommand{\nnum}[1]{\sepnum{.}{,}{}{#1}}
\definecolor{bittersweet}{rgb}{1.0, 0.44, 0.37}

%% file: main.bbl

\begin{thebibliography}{58}


\ifx \showCODEN    \undefined \def \showCODEN     #1{\unskip}     \fi
\ifx \showISBNx    \undefined \def \showISBNx     #1{\unskip}     \fi
\ifx \showISBNxiii \undefined \def \showISBNxiii  #1{\unskip}     \fi
\ifx \showISSN     \undefined \def \showISSN      #1{\unskip}     \fi
\ifx \showLCCN     \undefined \def \showLCCN      #1{\unskip}     \fi
\ifx \shownote     \undefined \def \shownote      #1{#1}          \fi
\ifx \showarticletitle \undefined \def \showarticletitle #1{#1}   \fi
\ifx \showURL      \undefined \def \showURL       {\relax}        \fi
\providecommand\bibfield[2]{#2}
\providecommand\bibinfo[2]{#2}
\providecommand\natexlab[1]{#1}
\providecommand\showeprint[2][]{arXiv:#2}

\bibitem[ris(2025)]%
        {risc0-guest-optimization-guide}
 \bibinfo{year}{2025}\natexlab{}.
\newblock \bibinfo{title}{{RISC Zero Guest Optimization Guide}}.
\newblock
  \bibinfo{howpublished}{\url{https://dev.risczero.com/api/zkvm/optimization}}.
\newblock


\bibitem[a16z crypto(2024)]%
        {a16z-benchmarks}
\bibfield{author}{\bibinfo{person}{a16z crypto}.}
  \bibinfo{year}{2024}\natexlab{}.
\newblock \bibinfo{title}{{a16z crypto zkVM benchmarks}}.
\newblock
  \bibinfo{howpublished}{\url{https://github.com/a16z/zkvm-benchmarks}}.
\newblock


\bibitem[Aho et~al\mbox{.}(2006)]%
        {compiler-dragonbook}
\bibfield{author}{\bibinfo{person}{Alfred~V. Aho}, \bibinfo{person}{Monica~S.
  Lam}, \bibinfo{person}{Ravi Sethi}, {and} \bibinfo{person}{Jeffrey~D.
  Ullman}.} \bibinfo{year}{2006}\natexlab{}.
\newblock \bibinfo{booktitle}{\emph{Compilers: Principles, Techniques, and
  Tools (2nd Edition)}}.
\newblock \bibinfo{publisher}{Addison-Wesley Longman Publishing Co., Inc.},
  \bibinfo{address}{USA}.
\newblock
\showISBNx{0321486811}


\bibitem[Ames et~al\mbox{.}(2023)]%
        {ligero}
\bibfield{author}{\bibinfo{person}{Scott Ames}, \bibinfo{person}{Carmit Hazay},
  \bibinfo{person}{Yuval Ishai}, {and} \bibinfo{person}{Muthuramakrishnan
  Venkitasubramaniam}.} \bibinfo{year}{2023}\natexlab{}.
\newblock \showarticletitle{Ligero: lightweight sublinear arguments without a
  trusted setup}.
\newblock \bibinfo{journal}{\emph{Des. Codes Cryptogr.}} \bibinfo{volume}{91},
  \bibinfo{number}{11} (\bibinfo{year}{2023}), \bibinfo{pages}{3379--3424}.
\newblock
\href{https://doi.org/10.1007/S10623-023-01222-8}{doi:\nolinkurl{10.1007/S10623-023-01222-8}}


\bibitem[Ansel et~al\mbox{.}(2014)]%
        {opentuner}
\bibfield{author}{\bibinfo{person}{Jason Ansel}, \bibinfo{person}{Shoaib
  Kamil}, \bibinfo{person}{Kalyan Veeramachaneni}, \bibinfo{person}{Jonathan
  Ragan-Kelley}, \bibinfo{person}{Jeffrey Bosboom}, \bibinfo{person}{Una-May
  O'Reilly}, {and} \bibinfo{person}{Saman Amarasinghe}.}
  \bibinfo{year}{2014}\natexlab{}.
\newblock \showarticletitle{OpenTuner: An Extensible Framework for Program
  Autotuning}. In \bibinfo{booktitle}{\emph{International Conference on
  Parallel Architectures and Compilation Techniques (PACT)}}.
  \bibinfo{address}{Edmonton, Canada}.
\newblock
\urldef\tempurl%
\url{https://commit.csail.mit.edu/papers/2014/ansel-pact14-opentuner.pdf}
\showURL{%
\tempurl}


\bibitem[Arun et~al\mbox{.}(2023)]%
        {jolt}
\bibfield{author}{\bibinfo{person}{Arasu Arun}, \bibinfo{person}{Srinath
  Setty}, {and} \bibinfo{person}{Justin Thaler}.}
  \bibinfo{year}{2023}\natexlab{}.
\newblock \bibinfo{title}{Jolt: {SNARKs} for Virtual Machines via Lookups}.
\newblock \bibinfo{howpublished}{Cryptology {ePrint} Archive, Paper 2023/1217}.
\newblock
\urldef\tempurl%
\url{https://eprint.iacr.org/2023/1217}
\showURL{%
\tempurl}


\bibitem[Baldimtsi et~al\mbox{.}(2024)]%
        {zklogin}
\bibfield{author}{\bibinfo{person}{Foteini Baldimtsi},
  \bibinfo{person}{Konstantinos~Kryptos Chalkias}, \bibinfo{person}{Yan Ji},
  \bibinfo{person}{Jonas Lindstr{\o}m}, \bibinfo{person}{Deepak Maram},
  \bibinfo{person}{Ben Riva}, \bibinfo{person}{Arnab Roy},
  \bibinfo{person}{Mahdi Sedaghat}, {and} \bibinfo{person}{Joy Wang}.}
  \bibinfo{year}{2024}\natexlab{}.
\newblock \showarticletitle{zkLogin: Privacy-Preserving Blockchain
  Authentication with Existing Credentials}. In
  \bibinfo{booktitle}{\emph{Proceedings of the 2024 on {ACM} {SIGSAC}
  Conference on Computer and Communications Security, {CCS} 2024, Salt Lake
  City, UT, USA, October 14-18, 2024}},
  \bibfield{editor}{\bibinfo{person}{Bo~Luo}, \bibinfo{person}{Xiaojing Liao},
  \bibinfo{person}{Jun Xu}, \bibinfo{person}{Engin Kirda}, {and}
  \bibinfo{person}{David Lie}} (Eds.). \bibinfo{publisher}{{ACM}},
  \bibinfo{pages}{3182--3196}.
\newblock
\href{https://doi.org/10.1145/3658644.3690356}{doi:\nolinkurl{10.1145/3658644.3690356}}


\bibitem[{Barry Whitehat}(2018)]%
        {WhiteHat_roll_up_token}
\bibfield{author}{\bibinfo{person}{{Barry Whitehat}}.}
  \bibinfo{year}{2018}\natexlab{}.
\newblock \bibinfo{title}{Roll Up Token}.
\newblock
\urldef\tempurl%
\url{https://github.com/barryWhiteHat/roll\_up\_token}
\showURL{%
\tempurl}
\newblock
\shownote{Accessed: 2025-06-01}.


\bibitem[Bellés-Muñoz et~al\mbox{.}(2023)]%
        {circom}
\bibfield{author}{\bibinfo{person}{Marta Bellés-Muñoz},
  \bibinfo{person}{Miguel Isabel}, \bibinfo{person}{Jose~Luis Muñoz-Tapia},
  \bibinfo{person}{Albert Rubio}, {and} \bibinfo{person}{Jordi Baylina}.}
  \bibinfo{year}{2023}\natexlab{}.
\newblock \showarticletitle{Circom: A Circuit Description Language for Building
  Zero-Knowledge Applications}.
\newblock \bibinfo{journal}{\emph{IEEE Transactions on Dependable and Secure
  Computing}} \bibinfo{volume}{20}, \bibinfo{number}{6} (\bibinfo{year}{2023}),
  \bibinfo{pages}{4733--4751}.
\newblock
\href{https://doi.org/10.1109/TDSC.2022.3232813}{doi:\nolinkurl{10.1109/TDSC.2022.3232813}}


\bibitem[Ben{-}Sasson et~al\mbox{.}(2018)]%
        {STARK}
\bibfield{author}{\bibinfo{person}{Eli Ben{-}Sasson}, \bibinfo{person}{Iddo
  Bentov}, \bibinfo{person}{Yinon Horesh}, {and} \bibinfo{person}{Michael
  Riabzev}.} \bibinfo{year}{2018}\natexlab{}.
\newblock \showarticletitle{Scalable, transparent, and post-quantum secure
  computational integrity}.
\newblock \bibinfo{journal}{\emph{{IACR} Cryptol. ePrint Arch.}}
  (\bibinfo{year}{2018}), \bibinfo{pages}{46}.
\newblock
\urldef\tempurl%
\url{http://eprint.iacr.org/2018/046}
\showURL{%
\tempurl}


\bibitem[Ben~Sasson et~al\mbox{.}(2014)]%
        {zcash}
\bibfield{author}{\bibinfo{person}{Eli Ben~Sasson}, \bibinfo{person}{Alessandro
  Chiesa}, \bibinfo{person}{Christina Garman}, \bibinfo{person}{Matthew Green},
  \bibinfo{person}{Ian Miers}, \bibinfo{person}{Eran Tromer}, {and}
  \bibinfo{person}{Madars Virza}.} \bibinfo{year}{2014}\natexlab{}.
\newblock \showarticletitle{Zerocash: Decentralized Anonymous Payments from
  Bitcoin}. In \bibinfo{booktitle}{\emph{2014 IEEE Symposium on Security and
  Privacy}}. \bibinfo{pages}{459--474}.
\newblock
\href{https://doi.org/10.1109/SP.2014.36}{doi:\nolinkurl{10.1109/SP.2014.36}}


\bibitem[Ben-Sasson et~al\mbox{.}(2014)]%
        {SNARK-vonNeumann}
\bibfield{author}{\bibinfo{person}{Eli Ben-Sasson}, \bibinfo{person}{Alessandro
  Chiesa}, \bibinfo{person}{Eran Tromer}, {and} \bibinfo{person}{Madars
  Virza}.} \bibinfo{year}{2014}\natexlab{}.
\newblock \showarticletitle{Succinct {Non-Interactive} Zero Knowledge for a von
  Neumann Architecture}. In \bibinfo{booktitle}{\emph{23rd USENIX Security
  Symposium (USENIX Security 14)}}. \bibinfo{publisher}{USENIX Association},
  \bibinfo{address}{San Diego, CA}, \bibinfo{pages}{781--796}.
\newblock
\showISBNx{978-1-931971-15-7}
\urldef\tempurl%
\url{https://www.usenix.org/conference/usenixsecurity14/technical-sessions/presentation/ben-sasson}
\showURL{%
\tempurl}


\bibitem[Bruestle and Gafni(2023)]%
        {risc0}
\bibfield{author}{\bibinfo{person}{Jeremy Bruestle} {and} \bibinfo{person}{Paul
  Gafni}.} \bibinfo{year}{2023}\natexlab{}.
\newblock \showarticletitle{{RISC} {Zero} {zkVM}: {Scalable}, {Transparent}
  {Arguments} of {RISC}-{V} {Integrity}}.
\newblock  (\bibinfo{year}{2023}).
\newblock


\bibitem[Bucek et~al\mbox{.}(2018)]%
        {spec-cpu}
\bibfield{author}{\bibinfo{person}{James Bucek}, \bibinfo{person}{Klaus-Dieter
  Lange}, {and} \bibinfo{person}{J\'{o}akim v. Kistowski}.}
  \bibinfo{year}{2018}\natexlab{}.
\newblock \showarticletitle{SPEC CPU2017: Next-Generation Compute Benchmark}.
  In \bibinfo{booktitle}{\emph{Companion of the 2018 ACM/SPEC International
  Conference on Performance Engineering}} (Berlin, Germany)
  \emph{(\bibinfo{series}{ICPE '18})}. \bibinfo{publisher}{Association for
  Computing Machinery}, \bibinfo{address}{New York, NY, USA},
  \bibinfo{pages}{41--42}.
\newblock
\showISBNx{9781450356299}
\href{https://doi.org/10.1145/3185768.3185771}{doi:\nolinkurl{10.1145/3185768.3185771}}


\bibitem[Buterin(2025)]%
        {simplel1}
\bibfield{author}{\bibinfo{person}{Vitalik Buterin}.}
  \bibinfo{year}{2025}\natexlab{}.
\newblock \bibinfo{title}{Simple L1: Make a successful L1 by only doing the
  bare essentials}.
\newblock
\urldef\tempurl%
\url{https://vitalik.eth.limo/general/2025/05/03/simplel1.html}
\showURL{%
\tempurl}


\bibitem[Chaliasos et~al\mbox{.}(2024a)]%
        {sokzkps}
\bibfield{author}{\bibinfo{person}{Stefanos Chaliasos}, \bibinfo{person}{Jens
  Ernstberger}, \bibinfo{person}{David Theodore}, \bibinfo{person}{David Wong},
  \bibinfo{person}{Mohammad Jahanara}, {and} \bibinfo{person}{Benjamin
  Livshits}.} \bibinfo{year}{2024}\natexlab{a}.
\newblock \showarticletitle{SoK: What don't we know? Understanding Security
  Vulnerabilities in SNARKs}. In \bibinfo{booktitle}{\emph{33rd {USENIX}
  Security Symposium, {USENIX} Security 2024, Philadelphia, PA, USA, August
  14-16, 2024}}, \bibfield{editor}{\bibinfo{person}{Davide Balzarotti} {and}
  \bibinfo{person}{Wenyuan Xu}} (Eds.). \bibinfo{publisher}{{USENIX}
  Association}.
\newblock
\urldef\tempurl%
\url{https://www.usenix.org/conference/usenixsecurity24/presentation/chaliasos}
\showURL{%
\tempurl}


\bibitem[Chaliasos et~al\mbox{.}(2024b)]%
        {benchzkevms}
\bibfield{author}{\bibinfo{person}{Stefanos Chaliasos}, \bibinfo{person}{Itamar
  Reif}, \bibinfo{person}{Adri{\`{a}} Torralba{-}Agell}, \bibinfo{person}{Jens
  Ernstberger}, \bibinfo{person}{Assimakis Kattis}, {and}
  \bibinfo{person}{Benjamin Livshits}.} \bibinfo{year}{2024}\natexlab{b}.
\newblock \showarticletitle{Analyzing and Benchmarking {{ZK}}-{{R}}ollups}. In
  \bibinfo{booktitle}{\emph{6th Conference on Advances in Financial
  Technologies, {AFT} 2024, September 23-25, 2024, Vienna, Austria}}
  \emph{(\bibinfo{series}{LIPIcs}, Vol.~\bibinfo{volume}{316})},
  \bibfield{editor}{\bibinfo{person}{Rainer B{\"{o}}hme} {and}
  \bibinfo{person}{Lucianna Kiffer}} (Eds.). \bibinfo{publisher}{Schloss
  Dagstuhl - Leibniz-Zentrum f{\"{u}}r Informatik}, \bibinfo{pages}{6:1--6:24}.
\newblock
\href{https://doi.org/10.4230/LIPICS.AFT.2024.6}{doi:\nolinkurl{10.4230/LIPICS.AFT.2024.6}}


\bibitem[Division(2024)]%
        {npb-c}
\bibfield{author}{\bibinfo{person}{NASA Advanced Supercomputing~(NAS)
  Division}.} \bibinfo{year}{2024}\natexlab{}.
\newblock \bibinfo{title}{{NAS Parallel Benchmarks}}.
\newblock
\urldef\tempurl%
\url{https://www.nas.nasa.gov/software/npb.html}
\showURL{%
\tempurl}


\bibitem[Ernstberger et~al\mbox{.}(2024a)]%
        {zkbench}
\bibfield{author}{\bibinfo{person}{Jens Ernstberger}, \bibinfo{person}{Stefanos
  Chaliasos}, \bibinfo{person}{George Kadianakis}, \bibinfo{person}{Sebastian
  Steinhorst}, \bibinfo{person}{Philipp Jovanovic}, \bibinfo{person}{Arthur
  Gervais}, \bibinfo{person}{Benjamin Livshits}, {and} \bibinfo{person}{Michele
  Orr{\`{u}}}.} \bibinfo{year}{2024}\natexlab{a}.
\newblock \showarticletitle{zk-Bench: {A} Toolset for Comparative Evaluation
  and Performance Benchmarking of SNARKs}. In
  \bibinfo{booktitle}{\emph{Security and Cryptography for Networks - 14th
  International Conference, {SCN} 2024, Amalfi, Italy, September 11-13, 2024,
  Proceedings, Part {I}}} \emph{(\bibinfo{series}{Lecture Notes in Computer
  Science}, Vol.~\bibinfo{volume}{14973})},
  \bibfield{editor}{\bibinfo{person}{Clemente Galdi} {and}
  \bibinfo{person}{Duong~Hieu Phan}} (Eds.). \bibinfo{publisher}{Springer},
  \bibinfo{pages}{46--72}.
\newblock
\href{https://doi.org/10.1007/978-3-031-71070-4\_3}{doi:\nolinkurl{10.1007/978-3-031-71070-4\_3}}


\bibitem[Ernstberger et~al\mbox{.}(2024b)]%
        {doyouneed}
\bibfield{author}{\bibinfo{person}{Jens Ernstberger}, \bibinfo{person}{Stefanos
  Chaliasos}, \bibinfo{person}{Liyi Zhou}, \bibinfo{person}{Philipp Jovanovic},
  {and} \bibinfo{person}{Arthur Gervais}.} \bibinfo{year}{2024}\natexlab{b}.
\newblock \showarticletitle{Do You Need a Zero Knowledge Proof?}
\newblock \bibinfo{journal}{\emph{{IACR} Cryptol. ePrint Arch.}}
  (\bibinfo{year}{2024}), \bibinfo{pages}{50}.
\newblock
\urldef\tempurl%
\url{https://eprint.iacr.org/2024/050}
\showURL{%
\tempurl}


\bibitem[{Ethereum Foundation}(2025)]%
        {ethproofs}
\bibfield{author}{\bibinfo{person}{{Ethereum Foundation}}.}
  \bibinfo{year}{2025}\natexlab{}.
\newblock \bibinfo{title}{Ethproofs}.
\newblock
\urldef\tempurl%
\url{https://ethproofs.org/}
\showURL{%
\tempurl}
\newblock
\shownote{Accessed: 2025-08-12}.


\bibitem[Ferrer(2020)]%
        {polybench-rust}
\bibfield{author}{\bibinfo{person}{Joseph~Rafael Ferrer}.}
  \bibinfo{year}{2020}\natexlab{}.
\newblock \bibinfo{title}{polybench-rs}.
\newblock \bibinfo{howpublished}{\url{https://github.com/JRF63/polybench-rs}}.
\newblock


\bibitem[Fog(2022)]%
        {instruction-tables-traditional-architectures}
\bibfield{author}{\bibinfo{person}{Agner Fog}.}
  \bibinfo{year}{2022}\natexlab{}.
\newblock \bibinfo{booktitle}{\emph{Instruction Tables: Lists of instruction
  latencies, throughputs and micro-operation breakdowns for Intel, AMD, and VIA
  CPUs}}.
\newblock Technical University of Denmark.
\newblock
\urldef\tempurl%
\url{https://www.agner.org/optimize/instruction_tables.pdf}
\showURL{%
\tempurl}
\newblock
\shownote{Last updated: 2022-11-04}.


\bibitem[Frigo and Shelat(2024)]%
        {longfellow}
\bibfield{author}{\bibinfo{person}{Matteo Frigo} {and} \bibinfo{person}{Abhi
  Shelat}.} \bibinfo{year}{2024}\natexlab{}.
\newblock \showarticletitle{Anonymous credentials from {ECDSA}}.
\newblock \bibinfo{journal}{\emph{{IACR} Cryptol. ePrint Arch.}}
  (\bibinfo{year}{2024}), \bibinfo{pages}{2010}.
\newblock
\urldef\tempurl%
\url{https://eprint.iacr.org/2024/2010}
\showURL{%
\tempurl}


\bibitem[Gabizon et~al\mbox{.}(2019)]%
        {plonk}
\bibfield{author}{\bibinfo{person}{Ariel Gabizon}, \bibinfo{person}{Zachary~J.
  Williamson}, {and} \bibinfo{person}{Oana Ciobotaru}.}
  \bibinfo{year}{2019}\natexlab{}.
\newblock \showarticletitle{{PLONK:} Permutations over Lagrange-bases for
  Oecumenical Noninteractive arguments of Knowledge}.
\newblock \bibinfo{journal}{\emph{{IACR} Cryptol. ePrint Arch.}}
  (\bibinfo{year}{2019}), \bibinfo{pages}{953}.
\newblock
\urldef\tempurl%
\url{https://eprint.iacr.org/2019/953}
\showURL{%
\tempurl}


\bibitem[Goldberg et~al\mbox{.}(2021)]%
        {Goldberg2021CairoA}
\bibfield{author}{\bibinfo{person}{Lior Goldberg}, \bibinfo{person}{Shahar
  Papini}, {and} \bibinfo{person}{Michael Riabzev}.}
  \bibinfo{year}{2021}\natexlab{}.
\newblock \showarticletitle{Cairo - a Turing-complete STARK-friendly CPU
  architecture}.
\newblock \bibinfo{journal}{\emph{IACR Cryptol. ePrint Arch.}}
  \bibinfo{volume}{2021} (\bibinfo{year}{2021}), \bibinfo{pages}{1063}.
\newblock
\urldef\tempurl%
\url{https://api.semanticscholar.org/CorpusID:237263398}
\showURL{%
\tempurl}


\bibitem[Goldwasser et~al\mbox{.}(1985)]%
        {zkp}
\bibfield{author}{\bibinfo{person}{S Goldwasser}, \bibinfo{person}{S Micali},
  {and} \bibinfo{person}{C Rackoff}.} \bibinfo{year}{1985}\natexlab{}.
\newblock \showarticletitle{{The knowledge complexity of interactive
  proof-systems}}. In \bibinfo{booktitle}{\emph{Proceedings of the Seventeenth
  Annual ACM Symposium on Theory of Computing}} (Providence, Rhode Island, USA)
  \emph{(\bibinfo{series}{STOC '85})}. \bibinfo{publisher}{Association for
  Computing Machinery}, \bibinfo{address}{New York, NY, USA},
  \bibinfo{pages}{291–304}.
\newblock
\showISBNx{0897911512}
\href{https://doi.org/10.1145/22145.22178}{doi:\nolinkurl{10.1145/22145.22178}}


\bibitem[Groth(2016)]%
        {groth16}
\bibfield{author}{\bibinfo{person}{Jens Groth}.}
  \bibinfo{year}{2016}\natexlab{}.
\newblock \showarticletitle{On the Size of Pairing-Based Non-interactive
  Arguments}. In \bibinfo{booktitle}{\emph{Proceedings, Part II, of the 35th
  Annual International Conference on Advances in Cryptology --- EUROCRYPT 2016
  - Volume 9666}}. \bibinfo{publisher}{Springer-Verlag},
  \bibinfo{address}{Berlin, Heidelberg}, \bibinfo{pages}{305--326}.
\newblock
\showISBNx{9783662498958}


\bibitem[Guibas(2024)]%
        {sp1-benchmarks}
\bibfield{author}{\bibinfo{person}{John Guibas}.}
  \bibinfo{year}{2024}\natexlab{}.
\newblock \bibinfo{title}{{SP1 Benchmarks}}.
\newblock
\urldef\tempurl%
\url{https://blog.succinct.xyz/sp1-benchmarks-8-6-24/}
\showURL{%
\tempurl}
\newblock
\shownote{Accessed: 2025-08-15}.


\bibitem[Inc.(2024)]%
        {aligned-zkvm-bench}
\bibfield{author}{\bibinfo{person}{Aligned Inc.}}
  \bibinfo{year}{2024}\natexlab{}.
\newblock \bibinfo{title}{{zkVMs benchmarks}}.
\newblock
  \bibinfo{howpublished}{\url{https://github.com/yetanotherco/zkvm_benchmarks}}.
\newblock


\bibitem[KENDALL(1938)]%
        {kendall}
\bibfield{author}{\bibinfo{person}{M.~G. KENDALL}.}
  \bibinfo{year}{1938}\natexlab{}.
\newblock \showarticletitle{A new measure of rank correlation}.
\newblock \bibinfo{journal}{\emph{Biometrika}} \bibinfo{volume}{30},
  \bibinfo{number}{1-2} (\bibinfo{date}{06} \bibinfo{year}{1938}),
  \bibinfo{pages}{81--93}.
\newblock
\showISSN{0006-3444}
\href{https://doi.org/10.1093/biomet/30.1-2.81}{doi:\nolinkurl{10.1093/biomet/30.1-2.81}}
\showeprint{https://academic.oup.com/biomet/article-pdf/30/1-2/81/423380/30-1-2-81.pdf}


\bibitem[Kothapalli et~al\mbox{.}(2022)]%
        {nova}
\bibfield{author}{\bibinfo{person}{Abhiram Kothapalli},
  \bibinfo{person}{Srinath T.~V. Setty}, {and} \bibinfo{person}{Ioanna
  Tzialla}.} \bibinfo{year}{2022}\natexlab{}.
\newblock \showarticletitle{Nova: Recursive Zero-Knowledge Arguments from
  Folding Schemes}. In \bibinfo{booktitle}{\emph{Advances in Cryptology -
  {CRYPTO} 2022 - 42nd Annual International Cryptology Conference, {CRYPTO}
  2022, Santa Barbara, CA, USA, August 15-18, 2022, Proceedings, Part {IV}}}
  \emph{(\bibinfo{series}{Lecture Notes in Computer Science},
  Vol.~\bibinfo{volume}{13510})}, \bibfield{editor}{\bibinfo{person}{Yevgeniy
  Dodis} {and} \bibinfo{person}{Thomas Shrimpton}} (Eds.).
  \bibinfo{publisher}{Springer}, \bibinfo{pages}{359--388}.
\newblock
\href{https://doi.org/10.1007/978-3-031-15985-5\_13}{doi:\nolinkurl{10.1007/978-3-031-15985-5\_13}}


\bibitem[Labs(2024a)]%
        {succinct-reth}
\bibfield{author}{\bibinfo{person}{Succinct Labs}.}
  \bibinfo{year}{2024}\natexlab{a}.
\newblock \bibinfo{title}{{Reth Succinct Processor (RSP)}}.
\newblock \bibinfo{howpublished}{\url{https://github.com/succinctlabs/rsp}}.
\newblock


\bibitem[Labs(2024b)]%
        {sp1}
\bibfield{author}{\bibinfo{person}{Succinct Labs}.}
  \bibinfo{year}{2024}\natexlab{b}.
\newblock \showarticletitle{{SP1} {Technical} {Whitepaper}}.
\newblock  (\bibinfo{year}{2024}).
\newblock
\newblock
\shownote{Whitepaper}.


\bibitem[Labs(2024c)]%
        {sp1-bridge}
\bibfield{author}{\bibinfo{person}{Succinct Labs}.}
  \bibinfo{year}{2024}\natexlab{c}.
\newblock \bibinfo{title}{{Upgrading Blobstream to SP1}}.
\newblock \bibinfo{howpublished}{\url{https://blog.succinct.xyz/celestia-sp1}}.
\newblock
\urldef\tempurl%
\url{https://blog.succinct.xyz/celestia-sp1}
\showURL{%
\tempurl}
\newblock
\shownote{Accessed: 2025-07-15}.


\bibitem[Labs(2024d)]%
        {succinct-zkvm-perf}
\bibfield{author}{\bibinfo{person}{Succinct Labs}.}
  \bibinfo{year}{2024}\natexlab{d}.
\newblock \bibinfo{title}{zkvm-perf}.
\newblock
  \bibinfo{howpublished}{\url{https://github.com/succinctlabs/zkvm-perf}}.
\newblock


\bibitem[Labs(2025)]%
        {sp1-github}
\bibfield{author}{\bibinfo{person}{Succinct Labs}.}
  \bibinfo{year}{2025}\natexlab{}.
\newblock \bibinfo{title}{{SP1 GitHub}}.
\newblock \bibinfo{howpublished}{\url{https://github.com/succinctlabs/sp1}}.
\newblock
\newblock
\shownote{GitHub repository}.


\bibitem[Lattner and Adve(2004)]%
        {llvm}
\bibfield{author}{\bibinfo{person}{C. Lattner} {and} \bibinfo{person}{V.
  Adve}.} \bibinfo{year}{2004}\natexlab{}.
\newblock \showarticletitle{{LLVM: a compilation framework for lifelong program
  analysis \& transformation}}. In \bibinfo{booktitle}{\emph{International
  Symposium on Code Generation and Optimization, 2004. CGO 2004.}}
  \bibinfo{pages}{75--86}.
\newblock
\href{https://doi.org/10.1109/CGO.2004.1281665}{doi:\nolinkurl{10.1109/CGO.2004.1281665}}


\bibitem[Lavin et~al\mbox{.}(2024)]%
        {zksurvey}
\bibfield{author}{\bibinfo{person}{Ryan Lavin}, \bibinfo{person}{Xuekai Liu},
  \bibinfo{person}{Hardhik Mohanty}, \bibinfo{person}{Logan Norman},
  \bibinfo{person}{Giovanni Zaarour}, {and} \bibinfo{person}{Bhaskar
  Krishnamachari}.} \bibinfo{year}{2024}\natexlab{}.
\newblock \showarticletitle{A Survey on the Applications of Zero-Knowledge
  Proofs}.
\newblock \bibinfo{journal}{\emph{CoRR}}  \bibinfo{volume}{abs/2408.00243}
  (\bibinfo{year}{2024}).
\newblock
\href{https://doi.org/10.48550/ARXIV.2408.00243}{doi:\nolinkurl{10.48550/ARXIV.2408.00243}}
\showeprint[arXiv]{2408.00243}


\bibitem[Ma et~al\mbox{.}(2023)]%
        {gzkp}
\bibfield{author}{\bibinfo{person}{Weiliang Ma}, \bibinfo{person}{Qian Xiong},
  \bibinfo{person}{Xuanhua Shi}, \bibinfo{person}{Xiaosong Ma},
  \bibinfo{person}{Hai Jin}, \bibinfo{person}{Haozhao Kuang},
  \bibinfo{person}{Mingyu Gao}, \bibinfo{person}{Ye Zhang},
  \bibinfo{person}{Haichen Shen}, {and} \bibinfo{person}{Weifang Hu}.}
  \bibinfo{year}{2023}\natexlab{}.
\newblock \showarticletitle{{GZKP:} {A} {GPU} Accelerated Zero-Knowledge Proof
  System}. In \bibinfo{booktitle}{\emph{Proceedings of the 28th {ACM}
  International Conference on Architectural Support for Programming Languages
  and Operating Systems, Volume 2, {ASPLOS} 2023, Vancouver, BC, Canada, March
  25-29, 2023}}, \bibfield{editor}{\bibinfo{person}{Tor~M. Aamodt},
  \bibinfo{person}{Natalie D.~Enright Jerger}, {and}
  \bibinfo{person}{Michael~M. Swift}} (Eds.). \bibinfo{publisher}{{ACM}},
  \bibinfo{pages}{340--353}.
\newblock
\href{https://doi.org/10.1145/3575693.3575711}{doi:\nolinkurl{10.1145/3575693.3575711}}


\bibitem[Martins et~al\mbox{.}(2025)]%
        {npb-rust}
\bibfield{author}{\bibinfo{person}{Eduardo~M. Martins},
  \bibinfo{person}{Leonardo~G. Faé}, \bibinfo{person}{Renato~B. Hoffmann},
  \bibinfo{person}{Lucas~S. Bianchessi}, {and} \bibinfo{person}{Dalvan
  Griebler}.} \bibinfo{year}{2025}\natexlab{}.
\newblock \bibinfo{title}{{NPB-Rust: NAS Parallel Benchmarks in Rust}}.
\newblock
\showeprint[arxiv]{2502.15536}~[cs.DC]
\urldef\tempurl%
\url{https://arxiv.org/abs/2502.15536}
\showURL{%
\tempurl}


\bibitem[Massalin(1987)]%
        {superoptimization}
\bibfield{author}{\bibinfo{person}{Henry Massalin}.}
  \bibinfo{year}{1987}\natexlab{}.
\newblock \showarticletitle{Superoptimizer: a look at the smallest program}. In
  \bibinfo{booktitle}{\emph{Proceedings of the Second International Conference
  on Architectual Support for Programming Languages and Operating Systems}}
  (Palo Alto, California, USA) \emph{(\bibinfo{series}{ASPLOS II})}.
  \bibinfo{publisher}{Association for Computing Machinery},
  \bibinfo{address}{New York, NY, USA}, \bibinfo{pages}{122–126}.
\newblock
\showISBNx{0818608056}
\href{https://doi.org/10.1145/36206.36194}{doi:\nolinkurl{10.1145/36206.36194}}


\bibitem[Mutmainnah(2024)]%
        {risc0-rollup}
\bibfield{author}{\bibinfo{person}{Mashiat Mutmainnah}.}
  \bibinfo{year}{2024}\natexlab{}.
\newblock \bibinfo{title}{{How to leverage RISC Zero's zkVM to scale Bitcoin}}.
\newblock
\urldef\tempurl%
\url{https://risczero.com/blog/how-to-leverage-risc-zeros-zkvm-to-scale-bitcoin}
\showURL{%
\tempurl}


\bibitem[Parno et~al\mbox{.}(2013)]%
        {pinocchio}
\bibfield{author}{\bibinfo{person}{Bryan Parno}, \bibinfo{person}{Jon Howell},
  \bibinfo{person}{Craig Gentry}, {and} \bibinfo{person}{Mariana Raykova}.}
  \bibinfo{year}{2013}\natexlab{}.
\newblock \showarticletitle{Pinocchio: Nearly Practical Verifiable
  Computation}. In \bibinfo{booktitle}{\emph{2013 {IEEE} Symposium on Security
  and Privacy, {SP} 2013, Berkeley, CA, USA, May 19-22, 2013}}.
  \bibinfo{publisher}{{IEEE} Computer Society}, \bibinfo{pages}{238--252}.
\newblock
\href{https://doi.org/10.1109/SP.2013.47}{doi:\nolinkurl{10.1109/SP.2013.47}}


\bibitem[Pouchet(2010)]%
        {polybench-c}
\bibfield{author}{\bibinfo{person}{Louis-Noel Pouchet}.}
  \bibinfo{year}{2010}\natexlab{}.
\newblock \bibinfo{title}{PolyBench/C - the Polyhedral Benchmark suite}.
\newblock
  \bibinfo{howpublished}{\url{https://www.cs.colostate.edu/~pouchet/software/polybench/}}.
\newblock


\bibitem[Ren et~al\mbox{.}(2021)]%
        {bintuner}
\bibfield{author}{\bibinfo{person}{Xiaolei Ren}, \bibinfo{person}{Michael Ho},
  \bibinfo{person}{Jiang Ming}, \bibinfo{person}{Yu Lei}, {and}
  \bibinfo{person}{Li Li}.} \bibinfo{year}{2021}\natexlab{}.
\newblock \showarticletitle{Unleashing the hidden power of compiler
  optimization on binary code difference: An empirical study}. In
  \bibinfo{booktitle}{\emph{Proceedings of the 42nd ACM SIGPLAN International
  Conference on Programming Language Design and Implementation}}.
  \bibinfo{pages}{142--157}.
\newblock


\bibitem[Sasnauskas et~al\mbox{.}(2018)]%
        {souper}
\bibfield{author}{\bibinfo{person}{Raimondas Sasnauskas}, \bibinfo{person}{Yang
  Chen}, \bibinfo{person}{Peter Collingbourne}, \bibinfo{person}{Jeroen
  Ketema}, \bibinfo{person}{Gratian Lup}, \bibinfo{person}{Jubi Taneja}, {and}
  \bibinfo{person}{John Regehr}.} \bibinfo{year}{2018}\natexlab{}.
\newblock \bibinfo{title}{{Souper: A Synthesizing Superoptimizer}}.
\newblock
\showeprint[arxiv]{1711.04422}~[cs.PL]
\urldef\tempurl%
\url{https://arxiv.org/abs/1711.04422}
\showURL{%
\tempurl}


\bibitem[Schkufza et~al\mbox{.}(2012)]%
        {stoke}
\bibfield{author}{\bibinfo{person}{Eric Schkufza}, \bibinfo{person}{Rahul
  Sharma}, {and} \bibinfo{person}{Alex Aiken}.}
  \bibinfo{year}{2012}\natexlab{}.
\newblock \bibinfo{title}{{Stochastic Superoptimization}}.
\newblock
\showeprint[arxiv]{1211.0557}~[cs.PF]
\urldef\tempurl%
\url{https://arxiv.org/abs/1211.0557}
\showURL{%
\tempurl}


\bibitem[Setty et~al\mbox{.}(2023)]%
        {lasso}
\bibfield{author}{\bibinfo{person}{Srinath Setty}, \bibinfo{person}{Justin
  Thaler}, {and} \bibinfo{person}{Riad Wahby}.}
  \bibinfo{year}{2023}\natexlab{}.
\newblock \bibinfo{title}{{Unlocking the lookup singularity with Lasso}}.
\newblock \bibinfo{howpublished}{Cryptology {ePrint} Archive, Paper 2023/1216}.
\newblock
\urldef\tempurl%
\url{https://eprint.iacr.org/2023/1216}
\showURL{%
\tempurl}


\bibitem[{Succinct Labs}(2025)]%
        {sp1-hardware-acceleration}
\bibfield{author}{\bibinfo{person}{{Succinct Labs}}.}
  \bibinfo{year}{2025}\natexlab{}.
\newblock \bibinfo{title}{SP1 Documentation: Hardware Acceleration}.
\newblock
  \bibinfo{howpublished}{\url{https://docs.succinct.xyz/docs/sp1/generating-proofs/hardware-acceleration}}.
\newblock
\newblock
\shownote{Accessed: 2025-08-12}.


\bibitem[Thaler et~al\mbox{.}(2022)]%
        {thaler2022proofs}
\bibfield{author}{\bibinfo{person}{Justin Thaler} {et~al\mbox{.}}}
  \bibinfo{year}{2022}\natexlab{}.
\newblock \showarticletitle{Proofs, arguments, and zero-knowledge}.
\newblock \bibinfo{journal}{\emph{Foundations and Trends{\textregistered} in
  Privacy and Security}} \bibinfo{volume}{4}, \bibinfo{number}{2--4}
  (\bibinfo{year}{2022}).
\newblock


\bibitem[Theodoridis et~al\mbox{.}(2022)]%
        {optimal-inlining}
\bibfield{author}{\bibinfo{person}{Theodoros Theodoridis},
  \bibinfo{person}{Tobias Grosser}, {and} \bibinfo{person}{Zhendong Su}.}
  \bibinfo{year}{2022}\natexlab{}.
\newblock \showarticletitle{Understanding and exploiting optimal function
  inlining}. In \bibinfo{booktitle}{\emph{Proceedings of the 27th ACM
  International Conference on Architectural Support for Programming Languages
  and Operating Systems}} (Lausanne, Switzerland)
  \emph{(\bibinfo{series}{ASPLOS '22})}. \bibinfo{publisher}{Association for
  Computing Machinery}, \bibinfo{address}{New York, NY, USA},
  \bibinfo{pages}{977–989}.
\newblock
\showISBNx{9781450392051}
\href{https://doi.org/10.1145/3503222.3507744}{doi:\nolinkurl{10.1145/3503222.3507744}}


\bibitem[Thomas et~al\mbox{.}(2025)]%
        {valida}
\bibfield{author}{\bibinfo{person}{Morgan Thomas}, \bibinfo{person}{Mamy
  Ratsimbazafy}, \bibinfo{person}{Marcin Bugaj}, \bibinfo{person}{Lewis
  Revill}, \bibinfo{person}{Carlo Modica}, \bibinfo{person}{Sebastian Schmidt},
  \bibinfo{person}{Ventali Tan}, \bibinfo{person}{Daniel Lubarov},
  \bibinfo{person}{Max Gillett}, {and} \bibinfo{person}{Wei Dai}.}
  \bibinfo{year}{2025}\natexlab{}.
\newblock \showarticletitle{Valida ISA Spec, version 1.0: A zk-Optimized
  Instruction Set Architecture}.
\newblock \bibinfo{journal}{\emph{arXiv preprint arXiv:2505.08114}}
  (\bibinfo{year}{2025}).
\newblock


\bibitem[Torczon and Cooper(2007)]%
        {compiler-engineering}
\bibfield{author}{\bibinfo{person}{Linda Torczon} {and} \bibinfo{person}{Keith
  Cooper}.} \bibinfo{year}{2007}\natexlab{}.
\newblock \bibinfo{booktitle}{\emph{Engineering A Compiler}
  (\bibinfo{edition}{2nd} ed.)}.
\newblock \bibinfo{publisher}{Morgan Kaufmann Publishers Inc.},
  \bibinfo{address}{San Francisco, CA, USA}.
\newblock
\showISBNx{012088478X}


\bibitem[{Verified zkEVM Initiative}(2025)]%
        {verifiedzkevm}
\bibfield{author}{\bibinfo{person}{{Verified zkEVM Initiative}}.}
  \bibinfo{year}{2025}\natexlab{}.
\newblock \bibinfo{title}{Verified zkEVM}.
\newblock
\urldef\tempurl%
\url{https://verified-zkevm.org/}
\showURL{%
\tempurl}


\bibitem[Waterman et~al\mbox{.}(2016)]%
        {riscv}
\bibfield{author}{\bibinfo{person}{Andrew Waterman}, \bibinfo{person}{Yunsup
  Lee}, \bibinfo{person}{David~A. Patterson}, {and} \bibinfo{person}{Krste
  Asanović}.} \bibinfo{year}{2016}\natexlab{}.
\newblock \bibinfo{booktitle}{\emph{The RISC-V Instruction Set Manual, Volume
  I: User-Level ISA, Version 2.1}}.
\newblock \bibinfo{type}{{T}echnical {R}eport} UCB/EECS-2016-118.
\newblock
\urldef\tempurl%
\url{http://www2.eecs.berkeley.edu/Pubs/TechRpts/2016/EECS-2016-118.html}
\showURL{%
\tempurl}


\bibitem[Zero(2025a)]%
        {risc0-recursion}
\bibfield{author}{\bibinfo{person}{RISC Zero}.}
  \bibinfo{year}{2025}\natexlab{a}.
\newblock \bibinfo{title}{{RISC Zero - Recursive Proving}}.
\newblock \bibinfo{howpublished}{\url{https://dev.risczero.com/api/recursion}}.
\newblock


\bibitem[Zero(2025b)]%
        {risc0-github}
\bibfield{author}{\bibinfo{person}{RISC Zero}.}
  \bibinfo{year}{2025}\natexlab{b}.
\newblock \bibinfo{title}{{RISC Zero GitHub}}.
\newblock \bibinfo{howpublished}{\url{https://github.com/risc0/risc0}}.
\newblock
\newblock
\shownote{GitHub repository}.


\end{thebibliography}
